\documentclass[%
 reprint,
superscriptaddress,
 amsmath,amssymb,
 aps,
 prd,
]{revtex4-2}

\usepackage{graphicx}
\graphicspath{{./figures/}}
\usepackage{xcolor}
\usepackage{ulem}
\usepackage{dcolumn}
\usepackage{bm}
\usepackage{hyperref}
\hypersetup{colorlinks, linkcolor = [rgb]{0, 0, 0.5}, citecolor = [rgb]{0,0.0,0.5}, urlcolor = [rgb]{0,0.0,0.5}}
\usepackage[caption=false]{subfig}
\usepackage{siunitx}
\usepackage[capitalise]{cleveref}
\allowdisplaybreaks

\newcommand{\xb}{x_{\mbox{\tiny\!$B$}}}

\begin{document}


\title{Towards three-dimensional nucleon structures at the Electron-Ion Collider in China: A study of the Sivers function}

\author{Chunhua Zeng}
\email{zengchunhua@impcas.ac.cn}
\affiliation{Institute of Modern Physics, Chinese Academy of Sciences, Lanzhou, Gansu Province 730000, China}
\affiliation{Lanzhou University, Lanzhou, Gansu Province 730000, China}
\affiliation{University of Chinese Academy of Sciences, Beijing 100049, China}

\author{Tianbo Liu}
\email{liutb@sdu.edu.cn}
\affiliation{Key Laboratory of Particle Physics and Particle Irradiation (MOE), Institute of Frontier and Interdisciplinary Science, Shandong University, Qingdao, Shandong 266237, China}

\author{Peng Sun}
\email{pengsun@impcas.ac.cn}
\affiliation{Institute of Modern Physics, Chinese Academy of Sciences, Lanzhou, Gansu Province 730000, China}
\affiliation{Department of Physics and Institute of Theoretical Physics,
Nanjing Normal University, Nanjing, Jiangsu 210023, China}

\author{Yuxiang Zhao}
\email{yxzhao@impcas.ac.cn}
\affiliation{Institute of Modern Physics, Chinese Academy of Sciences, Lanzhou, Gansu Province 730000, China}
\affiliation{University of Chinese Academy of Sciences, Beijing 100049, China}
\affiliation{Key Laboratory of Quark and Lepton Physics (MOE) and Institute of Particle Physics, Central China Normal University, Wuhan 430079, China}


\begin{abstract}
Transverse momentum-dependent parton distribution functions (TMDs) provide three-dimensional imaging of the nucleon in momentum space. With its fundamental importance in understanding the spin structure of the nucleon, the precise measurement of TMDs is considered  one of the main physics topics of the proposed Electron-Ion Collider in China (EicC). In this paper, we investigate the impact of future semi-inclusive deep inelastic scattering (SIDIS) data from EicC on the extraction of TMDs. Taking the Sivers function as an example, we revisited the world SIDIS data, which serves as the input for the simulated EicC data. By performing a combined analysis of the world data and the EicC simulated data, we quantitatively demonstrate that the Sivers functions can be precisely determined for various quark flavors, especially for sea quarks, in the region, $x>0.005$, directly covered by the EicC.
\end{abstract}

\maketitle

\section{Introduction}
\label{sec:introduction}

Nucleons are the basic building blocks of more than 98\% of the visible matter in the Universe. 
Understanding the nucleon structure in terms of the underlying quarks and gluons degrees of freedom is a central issue in modern nuclear and particle physics.
Because of the nonperturbative property of quantum chromodynamics (QCD) at hadronic energy scales, it is still a challenging task to directly evaluate nucleon structures from first principles~\cite{Constantinou:2022yye}, although much progress has been made in recent years~\cite{Ji:2013dva,Ma:2014jla,Radyushkin:2017cyf}.
The nucleon spin structures have received great interest since the first measurement of polarized lepton-nucleon deep inelastic scattering, which found the quark spin only contributes a small fraction to the nucleon spin~\cite{EuropeanMuon:1987isl,EuropeanMuon:1989yki} and triggered the so-called {\it proton spin puzzle}~\cite{Aidala:2012mv,Ji:2020ena}.
The remaining part of the nucleon spin is nowadays attributed to the gluon angular momentum and the quark orbital angular momentum, though one may have different decomposition versions~\cite{Jaffe:1989jz,Ji:1996ek,Leader:2013jra,Wakamatsu:2014zza}.
After more than 30 years of efforts from both experimental and theoretical aspects, the quark spin contribution is relatively well determined~\cite{COMPASS:2005xxc,COMPASS:2006mhr,HERMES:2006jyl} and the gluon spin part is also started being known~\cite{Nocera:2014gqa,deFlorian:2014yva,Ethier:2017zbq,Zhou:2022wzm}.
However, we have very rare knowledge of the orbital motion of quarks and gluons, which requires three-dimensional imaging of the nucleon. 

Since quarks and gluons cannot be directly observed in modern detectors, most analyses of hadron involved high energy scatterings rely on the QCD factorization~\cite{Collins:1989gx,Collins:2011zzd}, which approximates the cross section by the convolution of perturbatively calculable short-distance scattering off partons, including quarks and gluons, and universal long-distance functions.
The measurement of partonic structures of the nucleon was pioneered by SLAC~\cite{Bloom:1969kc} via the inclusive lepton-nucleon deep inelastic scattering (DIS) process,

\vspace{-0.5cm}
\begin{align}
    e(l) + N(P) \to e(l') + X,
\end{align}
where only the scattered lepton in the final state is detected and $X$ stands for the undetected hadronic system. The labels $l$, $P$, and $l'$ in parentheses represent the four-momenta of corresponding particles. The large momentum transfer $q = l - l'$ with $Q^2 = -q^2$ mediated by a virtual photon (or some other gauge boson) provides a hard scale $Q\gg \Lambda_{\rm QCD}$, which localizes the probe to ``see'' the partons in the nucleon. The inclusive DIS cross section can be expressed as the convolution of the lepton-parton hard scattering cross section and the collinear parton distribution function (PDF) $f_q(x,\mu)$, which describes the density of finding a parton of flavor $q$ carrying a fraction $x$ of the longitudinal momentum of the parent nucleon with $\mu$ representing the factorization scale.

In semi-inclusive DIS (SIDIS) process,

\vspace{-0.5cm}
\begin{align}
    e(l) + N(P) \to e(l') + h(P_h) + X,
    \label{eq:sidis}
\end{align}
a final-state hadron is identified in coincidence with the scattered lepton. In addition to the hard scale $Q$, which localizes the probe as in the inclusive DIS, the transverse momentum $P_{h\perp}$ of the hadron defined in the virtual photon-nucleon frame provides an adjustable scale. 
When $P_{h\perp}$ is integrated or compatible with $Q$, the SIDIS process effectively becomes a single-scale problem and one can apply the collinear factorization to approximate the cross section as the convolution of the hard scattering cross section, the collinear PDF $f_q(x,\mu)$, and the collinear fragmentation function $D_{q\to h}(z,\mu)$, which describes the probability density of a parton of flavor $q$ fragmenting to the hadron carrying longitudinal momentum fraction $z$.
When $P_{h\perp} \ll Q$, the SIDIS is a two-scale process, which is sensitive to the intrinsic transverse momenta of partons, and one should apply the transverse momentum dependent (TMD) factorization to express the cross section as the convolution of the hard scattering cross section, the TMD PDF $f_q(x,k_\perp,\mu)$, and the TMD FF $D_{q\to h}(z,p_\perp,\mu)$, where $k_\perp$ and $p_\perp$ represent the transverse momenta of the parton with respect to the hadron. The SIDIS is one of the main processes to measure TMD PDFs and hence to provide three-dimensional partonic structures of the nucleon in the momentum space.

Taking the spin degrees of freedom into account, we can learn about much richer nucleon structures, including the correlation between parton transverse momentum and the spin of the parton or the nucleon. At the leading twist, one can define eight TMD PDFs for the nucleon and only three of them have collinear counterparts. The Sivers function $f_{1T}^\perp(x,k_\perp)$, as one of the eight leading-twist TMD PDFs, was originally introduced to explain transverse single spin asymmetries in high energy scatterings~\cite{Sivers:1989cc,Sivers:1990fh}. It reflects the correlation between quark transverse momentum and the transverse spin of the nucleon. 
The Sivers function, as well as the Boer-Mulders function~\cite{Boer:1997nt}, is a naively time-reversal odd function, and thus , was believed to vanish in QCD for a long time~\cite{Collins:1992kk}. Motivated by some model calculation, it is found that the nonvanishing Sivers function can arise from the final-state interaction in the SIDIS process~\cite{Brodsky:2002cx} or the initial-state interaction in the Drell-Yan (DY) process~\cite{Brodsky:2002rv}. This effect is formally from the nontrivial Wilson line, which connects the quark field operator at different spacetime points to ensure the color gauge invariance~\cite{Collins:2002kn}. It is further proven in perturbative QCD factorization that the quark Sivers function in the SIDIS process and the one in DY process only differ by a sign,

\vspace{-0.5cm}
\begin{align}
    f_{1T}^\perp(x,k_\perp) \Big|_{\rm SIDIS} = - f_{1T}^\perp(x,k_\perp)\Big|_{\rm DY},
\end{align}
which receives great interest. Precise measurement of the Sivers function from both processes will provide an important test of the perturbative QCD (pQCD).

During the last two decades, many SIDIS experiments have been carried out by HERMES~\cite{HERMES:2009lmz,HERMES:2010mmo,HERMES:2020ifk}, COMPASS~\cite{COMPASS:2008isr,COMPASS:2014bze,COMPASS:2016led}, and JLab~\cite{JeffersonLabHallA:2011ayy,JeffersonLabHallA:2014yxb}, for explorations of nucleon TMD PDFs. The measured sizable target transverse single spin asymmetry (SSA), usually referred to as the Sivers asymmetry, can be interpreted at the leading twist via the convolution of the Sivers function and the unpolarized TMD FF, and indicates a nonzero Sivers function. Although it is the mostly studied polarized TMD PDF, the extractions of the Sivers function from global analyses of the SSA data still have large uncertainties~\cite{Echevarria:2020hpy,Bury:2021sue}, especially for sea quarks. The sign change prediction was tested through the $\pi N$ DY process~\cite{COMPASS:2017jbv} and the $pp$ collision with $W/Z$ production process~\cite{STAR:2015vmv}. The data do favor a sign flip of the Sivers function~\cite{Anselmino:2016uie}, but the large uncertainties are not able to exclude nonflip cases and other possibilities. All these pioneering experiments have provided valuable data and revealed rich information about the Sivers function and more generally the three-dimensional structures of the nucleon. However, our current knowledge of the Sivers function is still far from a satisfactory level, especially for sea quarks. Neither the shape of the function is well determined, nor the sign is known for all flavors. 

We are nowadays at the stage toward the precise measurement of TMD PDFs. The multihall SIDIS program at the 12 GeV upgraded JLab aims at the measurement at a relatively large-$x$ region, sometimes referred to as the valence quark region. The Electron-Ion Collider (EIC)~\cite{Accardi:2012qut,AbdulKhalek:2021gbh} to be built at BNL has unique advantages to reach the small-$x$ region, down to about $10^{-4}$. The Electron-Ion Collider in China (EicC)~\cite{Anderle:2021wcy} was recently proposed as one of the next-generation colliders in the world. It is designed to deliver a $3.5\,\rm GeV$ election beam colliding with a $20\,\rm GeV$ polarized proton beam or a polarized $^3\rm He$ beam, as well as several other kinds of ion beams. The ion beam can be longitudinally or transversely polarized with 70\% polarization, and the electron beam is polarized with 80\% polarization.
The instantaneous luminosity of $ep$ collision can reach $2\times 10^{33}\,\rm cm^{-2}\, s^{-1}$. The precise measurement of nucleon three-dimensional structures, especially the sea quark distributions, is one of the main physics goals at EicC. 
The combination of the polarized proton beam and the effectively polarized neutron beam and the capability of particle identifications of pions and kaons allows a complete separation of all light quark flavors~\cite{Anderle:2021dpv}.
The kinematic coverage of EicC can fill the gap between JLab and EIC. The combination of the three projects is expected to provide complete three-dimensional imaging of the nucleon, varying from low scale to high scale and from large $x$ to small $x$.

In this paper, we take the Sivers function as an example to investigate the impact of the EicC three-dimensional nucleon spin structure program on the determination of TMD PDFs.
We perform a global analysis of existing world SIDIS SSA data including the TMD evolution as the baseline. The improvement of EicC is then estimated by adding simulated pseudodata of semi-inclusive charged pion and charged kaon productions from both $ep$ and $e ^3\rm He$ collisions. 
The paper is organized as follows. In Sec.~\ref{sec:theory}, we briefly summarize the theoretical framework to extract the Sivers function from SIDIS target transverse SSA data, leaving some detailed formulas in Appendix~\ref{sec:appendix}. The global analysis of world data and the EicC projection are presented in Sec.~\ref{sec:fit}, followed by the summary in Sec.~\ref{sec:summary}.

\section{Theoretical formalism}
\label{sec:theory}

\subsection{TMD factorization formula}
\label{section:formalism}

We consider the SIDIS process~\eqref{eq:sidis} on a transversely polarized nucleon. With the one-photon-exchange approximation, one can express the differential cross section as

\vspace{-0.5cm}
\begin{align}
\frac{d\sigma(S_\perp)}{d\xb dy dz_h dP_{h\perp}^2 d\phi_h d\phi_S}=\sigma_0\left[ F_{UU} + \epsilon_{\perp\alpha\beta} S_{\perp}^{\alpha}F_{UT}^{\beta} + ...  \right],
\end{align}
where 
\begin{eqnarray}
\sigma_0 = \frac{\alpha^2}{\xb y Q^2}\frac{1-y+\frac{1}{2}y^2+\frac{1}{4}\gamma^2y^2}{1+\gamma^2}\left(1+\frac{\gamma^2}{2\xb}\right),
\end{eqnarray}
and $S_\perp^\alpha$ represents the transverse polarization of the nucleon.
As commonly used in the SIDIS process, we define the kinematic variables

\vspace{-0.5cm}
\begin{align*}
    Q^2 &= -q^2 = -(l-l')^2,
    \\
    \xb &= \frac{Q^2}{2P\cdot q},
    \quad
    y = \frac{P\cdot q}{P\cdot l},
    \quad
    z_h = \frac{P\cdot P_h}{P\cdot q},
    \\
    \gamma &= \frac{2\xb M}{Q} = \frac{MQ}{P\cdot q},
\end{align*}
where $l$ is the incoming lepton momentum, $l'$ is the outgoing lepton momentum, $P$ is the incoming nucleon momentum, $P_h$ is the detected outgoing hadron momentum, and $M$ is the nucleon mass. The transverse antisymmetric tensor is

\begin{figure}[htp]
    \centering
    \includegraphics[width=0.98\columnwidth]{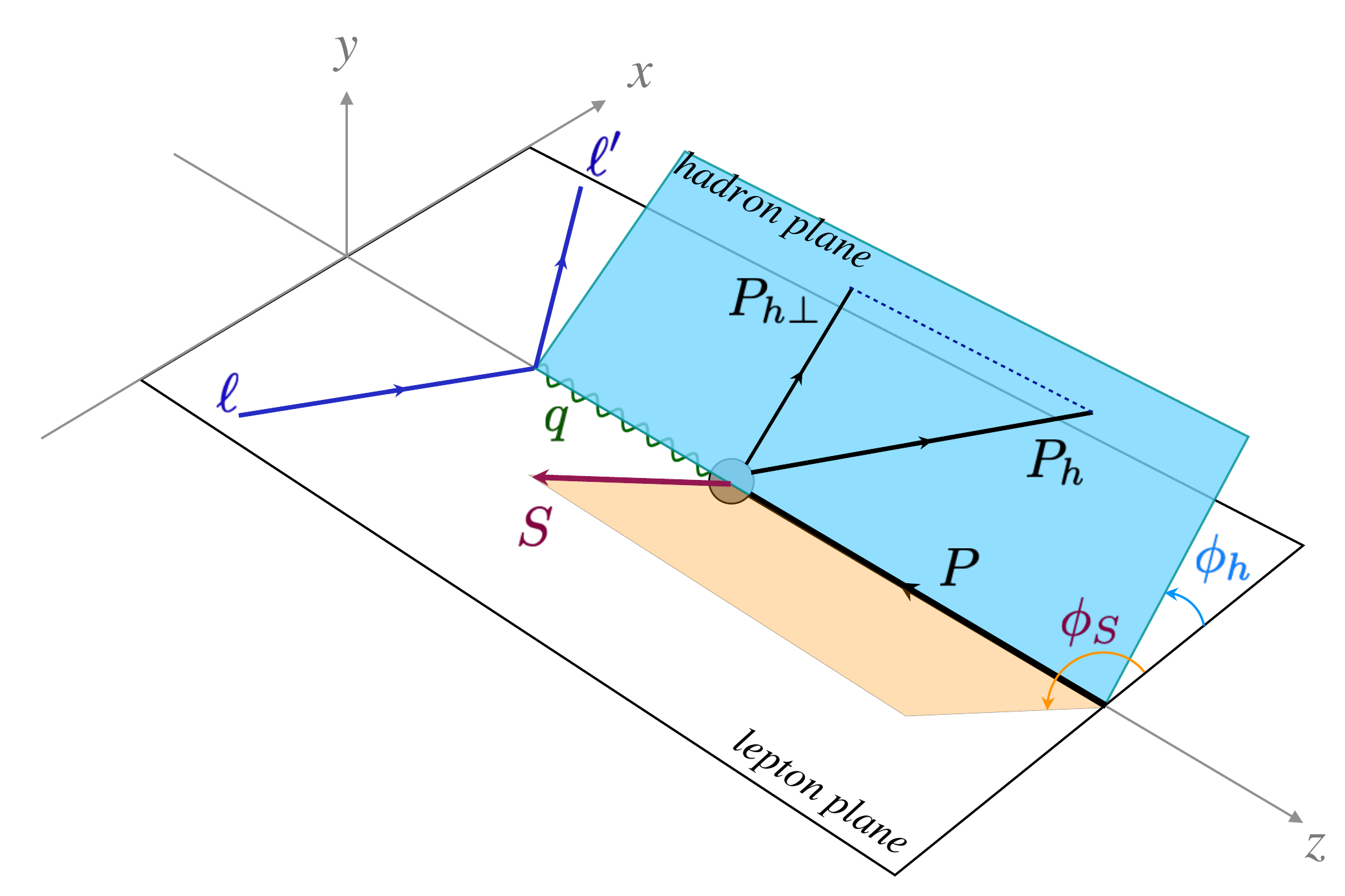}
    \caption{The Trento conventions of SIDIS kinematic variables.}
    \label{fig:frame}
\end{figure}

\vspace{-0.5cm}
\begin{align}
    \epsilon_{\perp}^{\mu \nu}&=\epsilon^{\mu \nu \rho \sigma} \frac{P_{\rho} q_{\sigma}}{P \cdot q \sqrt{1+\gamma^{2}}},
\end{align}
with the convention $\epsilon^{0123}=1$.
Following the Trento convention~\cite{Bacchetta:2004jz}, we define the hadron transverse momentum $P_{h\perp}$ and azimuthal angles in the virtual photon-nucleon frame, as illustrated in Fig.~\ref{fig:frame}. The $\phi_h$ is the angle from the lepton plane to the hadron plane and the $\phi_S$ is the angle from the lepton plane to the transverse spin $S_\perp$ of the nucleon. These variables can also be expressed in Lorentz invariant forms as

\vspace{-0.5cm}
\begin{align*}
    P_{h\perp} &= \sqrt{-g_\perp^{\mu\nu} P_{h\mu} P_{h\nu}},\\
    l_{\perp} &= \sqrt{-g_\perp^{\mu\nu} l_{\mu} l_{\nu}},\\
    \cos \phi_{h} &=-\frac{l_{\mu} P_{h \nu} g_{\perp}^{\mu \nu}}{l_{\perp} P_{h \perp}},
    \quad
    \sin \phi_{h}=-\frac{l_{\mu} P_{h \nu} \epsilon_{\perp}^{\mu \nu}}{l_{\perp} P_{h \perp}},
    \\
    \cos \phi_{S} &=-\frac{l_{\mu} S_{\perp\nu} g_{\perp}^{\mu \nu}}{l_{\perp} S_{\perp}},
    \quad
    \sin \phi_{S}=-\frac{l_{\mu} S_{\perp\nu} \epsilon_{\perp}^{\mu \nu}}{l_{\perp} S_{\perp}},
\end{align*}
where

\vspace{-0.5cm}
\begin{align}
    g_{\perp}^{\mu \nu}&=g^{\mu \nu}-\frac{q^{\mu} P^{\nu}+P^{\mu} q^{\nu}}{P \cdot q\left(1+\gamma^{2}\right)}+\frac{\gamma^{2}}{1+\gamma^{2}}\left(\frac{q^{\mu} q^{\nu}}{Q^{2}}-\frac{P^{\mu} P^{\nu}}{M^{2}}\right).
\end{align}

The structure functions $F_{UU}$ and $F_{UT}^\beta$ are functions of $\xb$, $z_h$, $P_{h\perp}$, and $Q$. For the convenience to apply the QCD factorization, one usually expresses the structure functions in the ``W+Y'' formalism~\cite{Collins:1984kg},

\vspace{-0.5cm}
\begin{align}
F_{UU}(\xb,z_h,P_{h\perp},Q)&=W_{UU}(\xb,z_h,P_{h\perp},Q) \notag\\
&\quad +Y_{UU}(\xb,z_h,P_{h\perp},Q), \\
F_{UT}^{\beta}(\xb,z_h,P_{h\perp},Q)&=W_{UT}^{\beta}(\xb,z_h,P_{h\perp},Q) \notag\\
&\quad +Y^{\beta}_{UT}(\xb,z_h,P_{h\perp},Q).
\end{align}
The ``W'' term dominates the cross section in the small $P_{hT}$ region, where the TMD factorization works, and the ``Y'' term provides a matching to the collinear factorization region at large $P_{hT}$. In this study, we restrict to the small $P_{hT}$ region and neglect the ``Y'' term. The resummation formula of the ``W'' term is derived in the impact parameter space via a Fourier transform,

\vspace{-0.5cm}
\begin{align}
W_{UU}(\xb,z_h,P_{h\perp},Q)&=\int\frac{d^2\vec{b}}{(2\pi)^2}
e^{-i\vec{q}_\perp\cdot
\vec{b}}\widetilde{W}_{UU}(\xb,z_h,b,Q),\\
W_{UT}^{\beta}(\xb,z_h,P_{h\perp},Q)&=\int\frac{d^2\vec{b}}{(2\pi)^2}
e^{-i\vec{q}_\perp\cdot
\vec{b}}\widetilde{W}_{UT}^{\beta}(\xb,z_h,b,Q), 
\end{align}
where $q_\perp = P_{h\perp}/z_h$ can be understood as the transverse momentum of the virtual photon in the nucleon-hadron back-to-back frame. 

According to the TMD factorization, one has the factorized expressions as

\vspace{-0.5cm}
\begin{align}
&\widetilde{W}_{UU}(\xb,z_h,b,Q) \notag\\ 
&=H(\mu,Q) \otimes f_1(x,b;\mu,\zeta_1) \otimes D_1(z,b;\mu,\zeta_2), \notag\\
&\widetilde{W}_{UT}^{\beta}(\xb,z_h,b,Q) \notag\\
&=H(\mu,Q) \otimes b^{\beta} M f^{\bot}_{1T}(x,b;\mu,\zeta_1) \otimes D_1(z,b;\mu,\zeta_2),
\label{resum}
\end{align}
with corrections suppressed by powers of $q_\perp/Q$.
The $H(\mu,Q)$ is the hard factor describing the short-distance scattering. In the calculation of the nucleon transverse SSA, the hard factors cancel out between the denominator and the numerator for fixed $\mu$ and $Q$, and hence we do not consider its effect in this work. 
$f_1(x,b;\mu,\zeta)$, $f^{\perp}_{1T}(x,b;\mu,\zeta)$ and $D_1(z,b;\mu,\zeta)$ are, respectively, the unpolarized TMD PDF, the Sivers function, and the unpolarized TMD FF in the $b$ space with $x$ representing the longitudinal momentum fraction carried by the active quark inside the nucleon and $z$ representing the longitudinal momentum fraction carried by the hadron from the fragmented quark. They are related to the corresponding functions in transverse momentum space via a Fourier transform,

\vspace{-0.5cm}
\begin{eqnarray}
f_1(x,k_\perp;\mu,\zeta)&=&\int_0^\infty \frac{db\, b}{2\pi} J_0(bk_\perp) f_1(x,b;\mu,\zeta)\notag \ ,\\
\frac{k_\perp}{M} f^{\perp}_{1T}(x,k_\perp;\mu,\zeta)&=&\int_0^\infty \frac{db\, b^2M}{2\pi} J_1(bk_\perp) f^{\perp}_{1T}(x,b;\mu,\zeta) \notag \ ,\\
D_1(z,p_\perp;\mu,\zeta)&=&\int_0^\infty \frac{db\, b}{2\pi} J_0(b p_\perp) D_1(z,b;\mu,\zeta)\ ,
\end{eqnarray}
where $J_{0,1}$ are the Bessel functions. The $\mu$ is the factorization scale, which one can choose arbitrarily.
The scales $\zeta_1$ and $\zeta_2$ in Eq.~\eqref{resum} are related to the invariant mass of virtual photon by $\zeta_1\zeta_2=Q^4$, and the ``W'' term only depends on their product $\zeta_1\zeta_2$. For convenience, we set the scales in the calculation as

\vspace{-0.5cm}
\begin{align}
 \mu^2 = \zeta_1 =\zeta_2=Q^2.
\end{align}
The convolutions over $x$ and $z$ in~\eqref{resum} are indicated by ``$\otimes$''. At the leading order, {\it i.e.}, the tree level, we have $\xb = x$ and $z_h = z$.

\subsection{Energy evolution and TMD resummation}
\label{sec:evolution}

In perturbation theory, one expands the factorized formula~\eqref{resum} in powers of the strong coupling constant $\alpha_s$. However, at each fixed order, {\it e.g.}, the $n$th order, the TMD functions $f_1(x,b;Q,Q^2)$, $f_{1T}^\perp(x,b;Q,Q^2)$, and $D_1(z,b;Q,Q^2)$ contain logarithmic enhanced terms of $[\alpha_s \ln^2(Qb)]^n$ and $[\alpha_s \ln(Qb)]^n$. In order to obtain a reliable prediction from perturbation theory, one needs to take into account the large logarithms of all orders. This can be achieved by evolving the scales $Q$ and $Q^2$ to $\mu_i \sim 1/b$ and $\zeta_i \sim 1/b^2$ and in the meantime resuming the large logarithms into an evolution factor $R[b;(\mu_i,\zeta_i)\to (Q,Q^2)]$.
Then one can formally relate the TMD functions at $(Q,Q^2)$ and $(\mu_i,\zeta_i)$ as

\vspace{-0.5cm}
\begin{align} \label{eq:6}
    &f_1(x,b;Q,Q^2)D_1(z,b;Q,Q^2)\notag\\
    =&R^2[b;(\mu_i,\zeta_i)\to (Q,Q^2)]f_1(x,b;\mu_i,\zeta_i)D_1(z,b;\mu_i,\zeta_i), \notag\\
    &f^{\bot}_{1T}(x,b;Q,Q^2)D_1(z,b;Q,Q^2)\notag\\
    =&R^2[b;(\mu_i,\zeta_i)\to (Q,Q^2)]f^{\bot}_{1T}(x,b;\mu_i,\zeta_i)D_1(z,b;\mu_i,\zeta_i).
\end{align}

The TMD evolution equations are given by

\vspace{-0.5cm}
\begin{align} 
    \frac{d}{d\ln\mu}f_1(x,b;\mu,\zeta)=\gamma_F(\mu,\zeta) f_1(x,b;\mu,\zeta)\label{evolv1},\\
    \frac{d}{d\ln\zeta}f_1(x,b;\mu,\zeta)=-\mathcal{D}(\mu,b) f_1(x,b;\mu,\zeta)\label{evolv2},
\end{align}
where $\gamma_F(\mu,\zeta)$ is the TMD anomalous dimension and $\mathcal{D}(\mu, b)$ is the rapidity anomalous dimension, with similar equations for  $f_{1T}^{\perp}(x,b;\mu,\zeta)$ and $D_1(z,b;\mu,\zeta)$.
By solving the equations above, one can express the factor $R[b;(\mu_i,\zeta_i)\to (Q,Q^2)]$ as a path integral from  $(\mu_i, \zeta_i)$ to $(Q, Q^2)$,

\vspace{-0.5cm}
\begin{align}\label{eq:evl}
    &R[b;(\mu_i,\zeta_i)\to (Q,Q^2)] \notag\\
    &= \exp\left[\int_{\cal P} \Big(\frac{\gamma_F(\mu, \zeta)}{\mu}d\mu-\frac{\mathcal{D}(\mu, b) }{\zeta}d\zeta  \Big)\right],
\end{align}
where the path $\cal P$ can be chosen arbitrarily because of the integrability condition

\vspace{-0.5cm}
\begin{align}
    \zeta \frac{d}{d\zeta} \gamma_F(\mu,\zeta) = -\mu \frac{d}{d\mu} {\cal D}(\mu,b) = -\Gamma_{\rm cusp}(\mu),
    \label{eq:cusp}
\end{align}
where $\Gamma_{\rm cusp}(\mu)$ is the cusp anomalous dimension.
Then the anomalous dimension $\gamma_F(\mu, \zeta)$ can be written as

\vspace{-0.5cm}
\begin{align}\label{eq:B2}
     \gamma_F(\mu, \zeta)=\Gamma_{\rm cusp}(\mu)\ln\Big(\frac{\mu^2}{\zeta}\Big)-\gamma_V(\mu), 
\end{align}
where $\gamma_V(\mu)$ is the finite part of the renormalization of the vector form factor. Expanding these factors in powers of the strong coupling constant,

\vspace{-0.5cm}
\begin{align}
\Gamma_{\rm cusp}(\mu)&=\sum_{n=0}^{\infty} a_s^{n+1} \Gamma_n,\\
\gamma_V(\mu)&=\sum_{n=1}^{\infty} a_s^n \gamma_n,
\end{align}
with $a_s=\alpha_s/(4\pi)$ introduced for convenience, one can calculate the coefficients $\Gamma_n$ and $\gamma_n$ perturbatively when $\mu\gg \Lambda_{\rm QCD}$, and up to two-loop level,

\vspace{-0.5cm}
\begin{align}
    \Gamma_0&=4C_F,\\
    \Gamma_1&=4C_F\Big[\big(\frac{67}{9}-\frac{\pi^2}{3}\big)C_A-\frac{20}{9}T_R N_f\Big],\\
    \gamma_1&=-6C_F,\\
    \gamma_2&=C_F^2(-3+4\pi^2-48\zeta_3)\notag\\
    &+C_F C_A\Big(-\frac{961}{27}-\frac{11\pi^2}{3}+52\zeta_3\Big)\notag\\
    &+C_F T_R N_f\Big(\frac{260}{27}+\frac{4\pi^2}{3}\Big),
\end{align}
where $C_F=4/3$, $C_A=3$, and $T_R=1/2$ are color factors of the $SU(3)$, $N_f=4$ is the number of active quark flavors, and $\zeta_3 \approx 1.202$ is the Ap\'ery's constant. 
Similarly, one can expand $\mathcal{D}(\mu, b)$ in powers of $a_s$ as,

\vspace{-0.5cm}
\begin{align}
\mathcal{D}_{\rm pert}(\mu, b)=\sum_{n=0}^{\infty}a_s^n  d_n(\textbf{L}_\mu),
\end{align}
which is only valid when $1/b \gg \Lambda_{\rm QCD}$, as indicated by the subscript ``pert''.
The $\textbf{L}_\mu$ is defined as

\vspace{-0.5cm}
\begin{align}
    \textbf{L}_\mu=\ln(\frac{\mu^2b^2}{4e^{-2\gamma_E}}),
\end{align}
with $\gamma_E$ the Euler-Mascheroni constant. Up to two-loop order, one has

\vspace{-0.5cm}
\begin{align}
    d_0(\textbf{L}_\mu)&=0,\\
    d_1(\textbf{L}_\mu)&=\frac{\Gamma_0}{2}\textbf{L}_\mu,\\
    d_2(\textbf{L}_\mu)&=\frac{\Gamma_0}{4}\beta_0\textbf{L}_\mu^2 + \frac{\Gamma_1}{2} \textbf{L}_\mu + d_2(0).
\end{align}
where 

\vspace{-0.5cm}
\begin{align}
     d_2(0)=C_F C_A\Big(\frac{404}{27}-14\zeta_3\Big)-\frac{112}{27}T_R N_f C_F
\end{align}
Following the treatment in~\cite{Echevarria:2012pw}, one can substitute $\mathcal{D}_{\rm pert}(\mu, b)$ into~\eqref{eq:cusp} to get a set of equations of $d_n({\bf L}_\mu)$. 
By solving these equations, one can sum the series of logarithmic terms $X^n$ for each $d_n(\textbf{L}_\mu)$, with $X=\beta_0 a_s\textbf{L}_\mu$, and obtain the resummed expression,

\vspace{-0.5cm}
\begin{align}
    &\mathcal{D}_{\rm resum}(\mu,b)=  
    -\frac{\Gamma_0}{2\beta_0}\ln(1-X)  \notag\\
    &+\frac{a_s}{2\beta_0(1-X)}\Big[-\frac{\beta_1\Gamma_0}{\beta_0}(\ln(1-X)+X)+\Gamma_1X    \Big]\notag\\
    &+\frac{a_s^2}{(1-X)^2}\Big[\frac{\Gamma_0\beta_1^2}{4\beta_0^3}(\ln^2(1-X)-X^2)   \notag\\
    &+\frac{\beta_1\Gamma_1}{4\beta_0^2}\big(X^2-2X-2\ln(1-X)\big) \notag\\
    &+\frac{\Gamma_0\beta_2}{4\beta_0^2}X^2-\frac{\Gamma_2}{4\beta_0}X(X-2) \notag\\
    &+ C_F C_A\Big(\frac{404}{27}-14\zeta_3\Big)-\frac{112}{27}T_R N_f C_F\Big].
\end{align}  
However, the perturbative expansion is invalid when $b \gtrsim 1/\Lambda_{\rm QCD}$, and hence one can only apply the above expression up to some $b$ value. For a smooth connection, one usually introduces the modified $b$ variable as

\vspace{-0.5cm}
\begin{align}\label{bs}
   b^*=\frac{b}{\sqrt{1+b^2/B^2_{\rm NP}}},
\end{align}
which serves as an effective truncation since $b^* < B_{\rm NP}$ holds for arbitrary large $b$ and $b^* \approx b$ for small $b$. 
Meanwhile, the large $b$ behavior of ${\cal D}(\mu,b)$ is modeled by a function $d_{\rm NP}$, which we adopt the form in Ref.~\cite{Bertone:2019nxa},

\vspace{-0.5cm}
\begin{align}
    d_{\rm NP}(b)=c_0bb^*,
\end{align}
which is linear function at large-$b$ region as suggested in Refs.~\cite{Tafat:2001in,Vladimirov:2020umg,Hautmann:2020cyp,Collins:2014jpa}.
We set the free parameters $B_{\rm NP}=1.93\,\rm GeV^{-1}$ and $c_0=0.0391\,\rm GeV^2$ as determined in Ref.~\cite{Scimemi:2019cmh} by fitting unpolarized SIDIS and DY data.
Finally, the $\mathcal{D}(\mu,b)$ is expressed as

\vspace{-0.5cm}
\begin{align}\label{Dterm}
    \mathcal{D}(\mu,b)= \mathcal{D}_{\rm resum}(\mu,b^*)+d_{\rm NP}(b) \ .
\end{align}

According to the $\zeta$ prescription~\cite{Scimemi:2019cmh}, by solving the equation,  

\vspace{-0.5cm}
\begin{align}
   \frac{d\ln \zeta_{\mu}(\mu, b) }{d \ln \mu^2 } = \frac{\gamma_F(\mu,\zeta_{\mu}(\mu, b))}{2\mathcal{D}(\mu, b)} \ ,
   \label{zeta_mu}
\end{align}
with the boundary conditions,

\vspace{-0.5cm}
\begin{align}
\mathcal{D}(\mu_0, b)=0  \ , 
\quad
\gamma_F(\mu_0,\zeta_{\mu}(\mu_0, b))=0 \ ,
\end{align}
one can find a special point $\zeta_{\mu}(Q, b)$, so that 

\vspace{-0.5cm}
\begin{align}\label{eq:RbQ}
    R[b;(\mu_i,\zeta_i)\to (Q,Q^2)]=\Big(\frac{Q^2}{\zeta_{\mu}(Q, b)}\Big)^{-\mathcal{D}(Q,b)} \ .
\end{align}
Then the Eq.~\eqref{zeta_mu} can be rewritten as

\vspace{-0.5cm}
\begin{align}
   \frac{d\ln \zeta_{\mu}( \mathcal{D}(\mu, b) , b) }{d \mathcal{D}(\mu, b) } \Gamma_{\rm cusp}   = \frac{\gamma_F(\mu,\zeta_{\mu}(\mu, b))}{2\mathcal{D}(\mu, b)} \ .
   \label{zeta_mu2}
\end{align}
Using Eq.~\eqref{Dterm} as an input, one can obtain a solution that is independent of the form of $\mathcal{D}(\mu, b)$ as

\vspace{-0.5cm}
\begin{align}
    \zeta^{\rm exact}_\mu(\mu,b)&=\mu^2e^{-g(\mu,b)/\mathcal{D}(\mu,b)} \ .
    \label{eq:exact}
\end{align}
Up to two-loop order, $g(\mu,b)$ can be written as

\vspace{-0.5cm}
\begin{align}
    &g(\mu,b) = 
    \frac{1}{a_s}\frac{\Gamma_0}{2\beta_0^2}\Bigg\{ e^{-p}-1+p
    +a_s\bigg[\frac{\beta_1}{\beta_0}\big(e^{-p}-1+p-\frac{p^2}{2}\big)  \notag\\
    &-\frac{\Gamma_1}{\Gamma_0}(e^{-p}-1+p)+\frac{\beta_0\gamma_1}{\Gamma_0}p \bigg] 
    +a_s^2\bigg[\Big(\frac{\Gamma_1^2}{\Gamma_0^2} -\frac{\Gamma_2}{\Gamma_0}\Big)(\cosh p -1)
    \notag\\
    &+\Big(\frac{\beta_1\Gamma_1}{\beta_0\Gamma_0}-\frac{\beta_2}{\beta_0}\Big)(\sinh p -p)
    +\Big(\frac{\beta_0\gamma_2}{\Gamma_0}-\frac{\beta_0\gamma_1\Gamma_1}{\Gamma_0^2}\Big)(e^p-1)\bigg]\Bigg\} \ ,
\end{align}
where

\vspace{-0.5cm}
\begin{align}
    p=\frac{2\beta_0\mathcal{D}(\mu,b)}{\Gamma_0} \ .
\end{align}
At extremely small $b$ region, there could be numerical difficulties for the solution~\eqref{eq:exact} to reduce to match the pQCD prediction~\cite{Scimemi:2019cmh}.  
On the other hand, $\mathcal{D}(\mu,b)$ will reduce to $\mathcal{D}_{\rm pert}(Q,b)$ since it is determined by pQCD in such a region. Instead of ${\cal D}(\mu, b)$, one use $\mathcal{D}_{\rm pert}(Q,b)$ as an input to solve Eq.~\eqref{zeta_mu} and will obtain the solution,

\vspace{-0.5cm}
\begin{align}
    \zeta^{\rm pert}_{\mu}(\mu,b)&=\frac{2\mu e^{-\gamma_E}}{b}e^{-v(\mu,b)} \ ,
\end{align}
which is consistent with the pQCD result by construction.
Up to two-loop order, $v(\mu,b)$ can be written as

\vspace{-0.5cm}
\begin{align}
    &v(\mu,b)=\frac{\gamma_1}{\Gamma_0}+a_s\Big[\frac{\beta_0}{12}\textbf{L}_\mu^2+\frac{\gamma_2+d_2(0)}{\Gamma_0}-\frac{\gamma_1\Gamma_1}{\Gamma_0^2}\Big] \ .
\end{align}
To match the large-$b$ region solution $\zeta_\mu^{\rm exact}(\mu, b)$ and the small-$b$ region solution $\zeta_\mu^{\rm pert}(\mu, b)$, we express $\zeta_\mu(\mu, b)$ as~\cite{Scimemi:2019cmh}:

\vspace{-0.5cm}
\begin{align}
    \zeta_{\mu}(\mu,b) &=
    \zeta^{\rm pert}_\mu(\mu,b)e^{-b^2/B_{\rm NP}^2} \notag\\
    &\quad +\zeta^{\rm exact}_\mu(\mu,b)\Big(1-e^{-b^2/B_{\rm NP}^2}\Big) \ .
\end{align}

\subsection{Unpolarized TMD PDF and FF}

Taking the phenomenological ansatzes in Ref.~\cite{Scimemi:2019cmh}, one can express the unpolarized TMD PDF and FF as

\vspace{-0.5cm}
\begin{align}\label{eq:OTD}
    &f_{1,f\gets h}(x,b;\mu_i,\zeta_i)
    =\sum_{f'}\int_x^1\frac{dy}{y}C_{f\gets f'}(y,b,\mu_{\rm OPE}^{\rm PDF})\notag\\
    &\quad\quad\quad\quad
    \times f_{1,f'\gets h}\Big(\frac{x}{y},\mu_{\rm OPE}^{\rm PDF}\Big)
    f_{\rm NP}(x,b) ,\\
    &D_{1,f\to h}(z,b;\mu_i,\zeta_i)
    =\frac{1}{z^2}\sum_{f'}\int_z^1\frac{dy}{y}y^2 \mathbb{C}_{f\to f'}(y,b,\mu_{\rm OPE}^{\rm FF})\notag\\
    &\quad\quad\quad\quad
    \times d_{1,f'\to h}\Big(\frac{z}{y},\mu_{\rm OPE}^{\rm FF}\Big)D_{\rm NP}(z,b) ,
\end{align}
where $f_{\rm NP}(x,b)$ and $D_{\rm NP}(z,b)$ are nonperturbative functions, $f_{1,f'\gets h}(x,\mu)$ and $d_{1,f'\to h}(x,\mu)$ are collinear PDF and FF, 
and $C_{f\gets f'}(y,b,\mu)$ and $\mathbb{C}_{f\to f'}(y,b,\mu)$ are matching coefficients that can be calculated via pQCD.
In this work, we consider the $C(\mathbb{C})$ functions up to the one-loop order, and the expressions are provided in Appendix~\ref{sec:appendix}. 
The evolution of scales $\mu_{\rm OPE}^{\rm PDF}$ and $\mu_{\rm OPE}^{\rm FF}$ in the $\zeta$ prescription are independent of external parameters. Here we adopt the choices~\cite{Scimemi:2019cmh}

\vspace{-0.5cm}
\begin{align}
  \mu_{\rm OPE}^{\rm PDF}&=\frac{2e^{-\gamma_E}}{b}+2\,{\rm GeV} \ ,\\
  \mu_{\rm OPE}^{\rm FF}&=\frac{2e^{-\gamma_E}z}{b}+2\,\rm GeV \ ,
\end{align}
and  $2\,\rm GeV$ is a typical reference scale for PDFs and FFs.
Correspondingly, we take the parametrization of the nonperturbative functions $f_{\rm NP}(x,b)$ and $D_{\rm NP}(z,b)$ as~ \cite{Scimemi:2019cmh}

\vspace{-0.5cm}
\begin{align}
    f_{\rm NP}(x,b)
    &=\exp\Big[-\frac{\lambda_1(1-x)+\lambda_2x+x(1-x)\lambda_5}{\sqrt{1+\lambda_3x^{\lambda_4}b^2}}b^2   \Big] ,\\
    D_{\rm NP}(z,b)&=\exp\Big[-\frac{\eta_1z+\eta_2(1-z)}{\sqrt{1+\eta_3(b/z)^2}}\frac{b^2}{z^2} \Big]
    \big(1+\eta_4\frac{b^2}{z^2} \big), 
\end{align}
where the parameters $\lambda$'s and $\eta$'s parameters $\lambda$ and $\eta$ are extracted from unpolarized SIDIS and Drell-Yan data at small transverse momentum. 
Their values in this study are listed in Table~\ref{Lam_ETA}.

\begin{table}[htp]
    \centering
    \caption{The values of the parameters in the parametrization of unpolarized TMD PDF and FF. Their units are in GeV$^2$ except for $\lambda_4$. }
    \begin{tabular}{ccccccccc}
    \hline\hline
    $\ \ \ \ \  \lambda_1$\ \ \ \ \  &\ \ \ \ \ $ \lambda_2$\ \ \ \ \       &\ \ \ \ \  $ \lambda_3$\ \ \ \ \       &\ \ \ \ \ $ \lambda_4$ \ \ \ \ \     & \ \ \ \ \ $\lambda_5$\ \ \ \ \  \\
  0.198     & 9.30    & 431   &   2.12 & -4.44 \\
 \hline \hline
 $ \eta_1$      &$ \eta_2$      & $ \eta_3$      &$ \eta_4$       \\
  0.260     & 0.476   & 0.478   &   0.483   \\
    \hline\hline
    \end{tabular}\label{Lam_ETA}
\end{table}

\subsection{ Sivers asymmetry}

The Sivers asymmetry in SIDIS is a transverse single spin asymmetry. It can be generally defined as

\vspace{-0.5cm}
\begin{align}
    A_{UT}^{\sin(\phi_h - \phi_S)}=&\frac{2\int d\phi_S d\phi_h[d\sigma^{\uparrow}-d\sigma^{\downarrow}]\sin(\phi_h-\phi_S)}{\int d\phi_S d\phi_h[d\sigma^{\uparrow}+d\sigma^{\downarrow}]},
\end{align}
where $\sigma^{\uparrow}$ and $\sigma^{\downarrow}$ represent the cross sections from transversely polarized nucleon. The azimuthal angles $\phi_S$ and $\phi_h$ are defined following the Trento convention as illustrated in~Fig.~\ref{fig:frame}.

\begin{table}[htp]
    \centering
    \caption{The precision of the  factors in powers of $\alpha_s$ in this work.}
    \label{tab:}
    \begin{tabular}{c|cccccccc}
    \hline\hline
    &\ \ $\Gamma_{cusp}$\ \ &\ \ $\gamma_V$\ \ &$\mathcal{D}_{resum}$\ \ &\ \ $\zeta_{\mu}^{pert}$\ \ &\ \ $\zeta_{\mu}^{exact}$\ \ &\ \ $C(\mathbb{C}$)\ \ \\
    $F_{UU}$& $\alpha_s^3$  & $\alpha_s^2$   &  $\alpha_s^2$  &  $\alpha_s^1$   & $\alpha_s^1$ &$\alpha_s^1$ \\
    $F_{UT}$& $\alpha_s^3$  & $\alpha_s^2$   &  $\alpha_s^2$  &  $\alpha_s^1$   & $\alpha_s^1$ &$\alpha_s^0$\\
    \hline\hline
    \end{tabular}\label{ORDERS}
\end{table}

Within the TMD formalism presented above, one can express the Sivers asymmetry $A_{UT}^{\sin(\phi_h - \phi_S)}$ as

\vspace{-0.5cm}
\begin{widetext}
\begin{align}\label{eq:AUTA}
     &A_{UT}^{\sin(\phi_h - \phi_S)}
     =
     -\frac{ M \sum_qe^2_q\int_0^\infty\frac{bdb}{2\pi}bJ_1(\frac{b|P_{hT}|}{z})R^2[b;(\mu_i,\zeta_i)\to (Q,Q^2)] f^{\bot}_{1T,q\gets h_1}(x,b)D_{1,q\to h_2}(z,b)}
    {\sum_qe^2_q\int_0^\infty\frac{bdb}{2\pi}J_0(\frac{b|P_{hT}|}{z})R^2[b;(\mu_i,\zeta_i)\to (Q,Q^2)] f_{1,q\gets h_1}(x,b)D_{1,q\to h_2}(z,b)}
    \ .
\end{align}
\end{widetext}
The precision for the perturbative calculation of the factors in powers of $\alpha_s$ in this
work is summarized in Table~\ref{ORDERS}.

\begin{table*}[tp]
\centering
\caption{World SIDIS data used in our analysis. The numbers in parentheses are the original number of
data points before applying $\delta$ cut.}
\label{table:world_data}
\begin{tabular*}{0.9\textwidth}{m{0.2\textwidth}m{0.15\textwidth}m{0.2\textwidth}m{0.15\textwidth}m{0.2\textwidth}}
\hline\hline
Dataset       & Target    & Beam      & Data points  & Reaction   \\ \hline
COMPASS~\cite{COMPASS:2008isr}  & $^{6}\text{LiD}$ & $160\,\rm GeV$ $\mu^+$  &10(41)  &$\mu^+d\to\mu^+\pi^+X $       \\
                                       &           & &&$\mu^+d\to\mu^+\pi^-X $          \\
                                    &           & &&$\mu^+d\to \mu^+K^+X  $          \\
                                           &           & &&$\mu^+d\to \mu^+K^-X  $          \\
                                             &           & &&$\mu^+d\to \mu^+K^0X  $         \\
\hline
COMPASS~\cite{COMPASS:2014bze}  & $\text{NH}_3  $  &$160\,\rm GeV$ $\mu^+$   &10(36)  &$\mu^+p\to\mu^+\pi^+X $     \\
                                             &           & &&$\mu^+p\to\mu^+\pi^-X $          \\
                                             &           & &&$\mu^+p\to \mu^+K^+X  $           \\
                                             &           & &&$\mu^+p\to \mu^+K^-X  $         \\
\hline
HERMES~\cite{HERMES:2020ifk}&  $\text{H}_2$  &$27.6\,\rm GeV$ $e^{\pm}$   &104(256)  &$e^{\pm}p\to e^{\pm}\pi^+X $          \\
                                             &           & &&$e^{\pm}p\to e^{\pm}\pi^-X  $           \\
                                             &           & &&$e^{\pm}p\to e^{\pm}K^+X  $            \\
                                             &           & &&$e^{\pm}p\to e^{\pm}K^-X  $            \\
\hline
JLab~\cite{JeffersonLabHallA:2011ayy}  & $^{3}\text{He} $ & $5.9\,\rm GeV$ $e^-$ &4(8)   &$e^-n\to e^-\pi^+X $       \\
                                          &           & &&$e^-n\to e^-\pi^-X $           \\
\hline
JLab~\cite{JeffersonLabHallA:2014yxb}  & $^{3}\text{He} $ & $5.9\,\rm GeV$ $e^-$ &2(5)   &$e^- {}^{3}{\rm He}\to e^-K^+X $         \\
                                             &           & &&$e^- {}^{3}{\rm He}\to e^-K^-X $             \\
\hline \hline
\end{tabular*}
\end{table*}

\section{Extract the Sivers function from the SIDIS data}
\label{sec:fit}

We parametrize the Sivers function of each quark flavor as

\vspace{-0.5cm}
\begin{align}
    f_{1T;q\gets p}^{\perp}(x,b)=N_q\frac{(1-x)^{\alpha_q}x^{\beta_q}(1+\epsilon_qx)}{n(\beta_q,\epsilon_q,\alpha_q)}\exp
\big(-r_{q}b^2\big),
\label{eq:param-form}
\end{align}
where $N_q$, $\alpha_q$, $\beta_q$, $\epsilon_q$, and $r_q$ are free parameters. The factor

\vspace{-0.5cm}
\begin{align*}
    n(\beta,\epsilon,\alpha)&=\frac{\Gamma(\alpha+1)(2+\alpha+\beta+\epsilon+\epsilon\beta)\Gamma(\beta+1)}{\Gamma(\beta+\alpha+3)}
\end{align*}
is introduced to reduce the correlation between the parameters controlling the shape and the parameter for the normalization.

To extract the Sivers function from the SIDIS SSA data, one also needs unpolarized TMD PDF $f_1(x,b)$ and unpolarized TMD FF $D_1(z,b)$, which we take from recent global analysis~\cite{Scimemi:2019cmh},
as mentioned in the previous section. For simplicity, we only consider nonvanishing Sivers functions for light flavors, {\it i.e.}, $u$, $d$, $s$, $\bar u$, $\bar d$, and $\bar s$, while heavy quark Sivers functions are all set to zero. We also assume the isospin symmetry to relate the Sivers function of the neutron and the Sivers function of the proton as

\vspace{-0.5cm}
\begin{align}
    f^{\perp}_{1T,u\gets n}(x,b)&=f^{\perp}_{1T,d\gets p}(x,b),\notag\\
    f^{\perp}_{1T,\bar{u}\gets n}(x,b)&=f^{\perp}_{1T,\bar{d}\gets p}(x,b),\notag\\
    f^{\perp}_{1T,d\gets n}(x,b)&=f^{\perp}_{1T,u\gets p}(x,b),\notag\\
    f^{\perp}_{1T,\bar{d}\gets n}(x,b)&=f^{\perp}_{1T,\bar{u}\gets p}(x,b),\notag\\
    f^{\perp}_{1T,s\gets n}(x,b)&=f^{\perp}_{1T,s\gets p}(x,b),\notag\\
    f^{\perp}_{1T,\bar{s}\gets n}(x,b)&=f^{\perp}_{1T,\bar{s}\gets p}(x,b).
\end{align}

Since a free neutron target is not available for SIDIS experiments, the polarized deuteron and polarized $^3\rm He$ are commonly used to obtain parton distributions in the neutron. As an approximation, we set the Sivers functions of the deuteron and the $^3\rm He$ via the weighted combination of the proton Sivers function and the neutron Sivers function. For the deuteron, the Sivers function is expressed as

\vspace{-0.5cm}
\begin{align}
    f^{\perp}_{1T,q\gets d}(x,b)&=\frac{P_{d}^n f^{\perp}_{1T,q\gets n}(x,b) + P_d^{p} f^{\perp}_{1T,q\gets p}(x,b)}{2},
\end{align}
where $P_d^n = P_d^p = 0.925$ are effective polarizations of the neutron and the proton in a polarized deuteron~\cite{Wiringa:1994wb}. Similarly, the Sivers function of the $^3\rm He$ is

\vspace{-0.5cm}
\begin{align}
    f^{\perp}_{1T,q\gets ^3\rm He}(x,b)&=\frac{P_{\rm He}^n f^{\perp}_{1T,q\gets n}(x,b) + 2P_{\rm He}^p f^{\perp}_{1T,q\gets p}(x,b)}{3},
\end{align}
where $P_{\rm He}^n = 0.86$ and $P_{\rm He}^p = -0.028$ are effective polarizations of the neutron and the proton in a polarized $^3\rm He$~\cite{Friar:1990vx}. Here we have followed the convention to normalize the distribution to per nucleon. 

This parametrization setup is applied in the following for both the fit to world SIDIS data and the fit to EicC pseudodata.

\subsection{Fit to world SIDIS data}
\label{section:fitworld}

We first perform the fit to existing SIDIS target transverse SSA data from HERMES~\cite{HERMES:2020ifk}, COMPASS~\cite{COMPASS:2008isr,COMPASS:2014bze}, and JLab~\cite{JeffersonLabHallA:2011ayy,JeffersonLabHallA:2014yxb}. For the COMPASS measurements, we only include the analyzed data with particle identifications of the final state hadron for the purpose of flavor separation, although the results without hadron particle identifications for the same datasets have also been published~\cite{COMPASS:2016led}.
For the validity of the TMD factorization, only small transverse momentum data with $\delta = |P_{h\perp}|/(z Q) < 0.5$ are selected. After applying this cut, there are 130 data points, as summarized in Table~\ref{table:world_data}. These data are all from fixed-target experiments with various beam energies: $160\,\rm GeV$ muon beam for COMPASS, $27.6\,\rm GeV$ electron/positron beam for HERMES, and $5.9\,\rm GeV$ electron beam for JLab.

Since the existing world data are not precise enough to constrain so many parameters introduced in~\eqref{eq:param-form} for all light flavors, we reduce the number of parameters by setting $s$ and $\bar{s}$ quarks Sivers function to $0$ and imposing the following assumptions for $u, d, \bar{u}, \bar{d}$ quarks:
\begin{equation}
    \begin{split}
    &r_{\bar{u}}=r_{\bar{d}}=r_{\rm sea},
    \quad
    \beta_{\bar{u}}=\beta_{\bar{d}}=\beta_{\rm sea},\\
    &\alpha_u=\alpha_d=3,
    \quad
    \alpha_{\bar{u}}=\alpha_{\bar{d}}=5,
    \quad
    \epsilon_{\bar u}= \epsilon_{\bar d} =  0.
    \end{split}
\label{eq:param-constraints}
\end{equation}
Then we have $12$ free parameters,  as listed in Table~\ref{tab:world_params}. To estimate the uncertainty, we randomly shift the central values of the data points by Gaussian distributions with the Gaussian widths to be the experimental uncertainties 
and then perform the fit to each smeared dataset. By repeating this procedure, 100 replicas are created for calculations of central values and uncertainties of physical
quantities related to the Sivers function. The total $\chi^2/N$ of the fit as well as its value
for various experimental datasets are listed in table \ref{table:world_chi2}. Here, $N$ is the number of experimental data points. 
The details of the calculation are explained
in Appendix \ref{sec:appendix_2}.

\begin{figure*}[htp]
    \centering
    \includegraphics[width=0.4\textwidth]{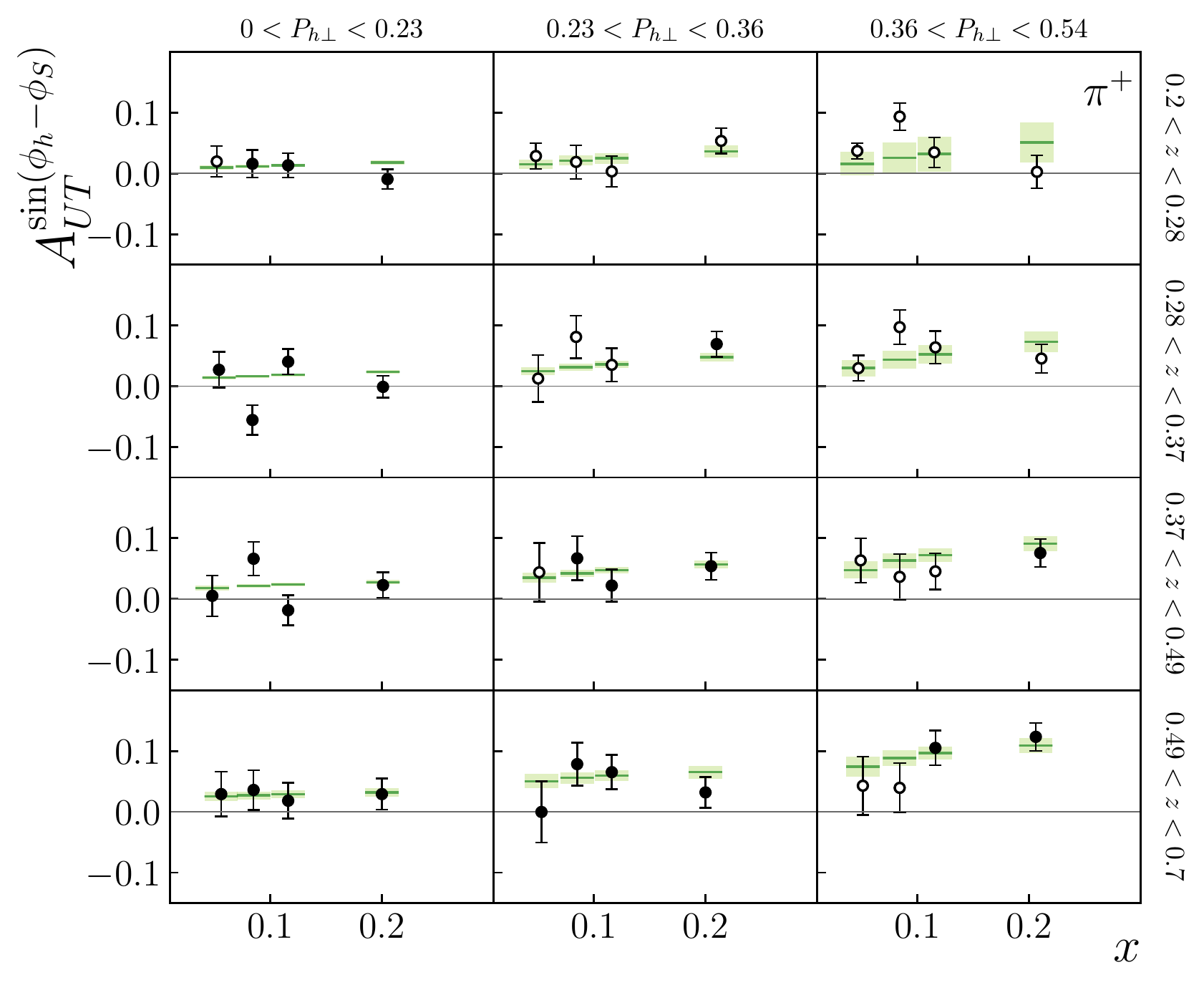}
    \includegraphics[width=0.4\textwidth]{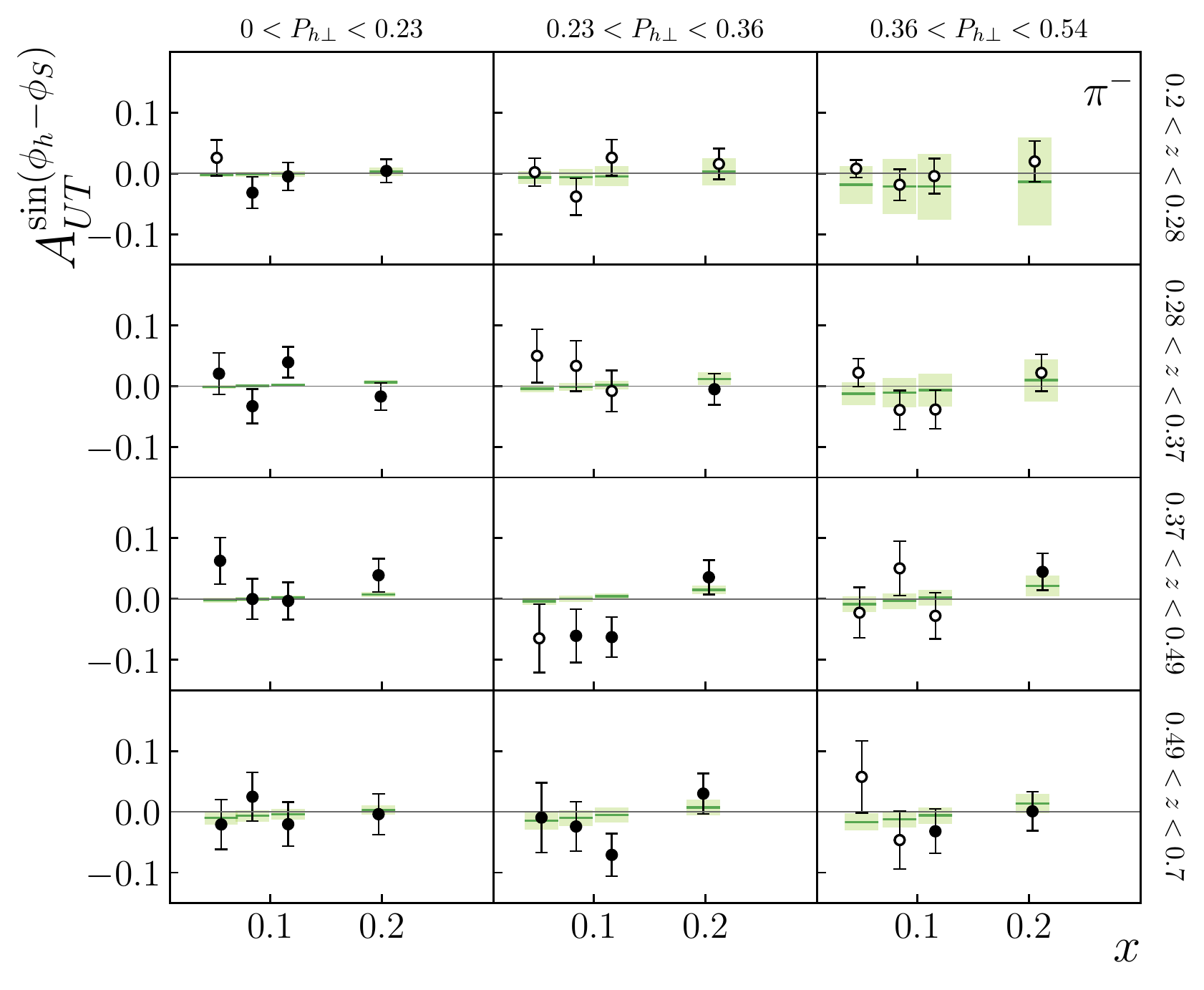}
    \includegraphics[width=0.4\textwidth]{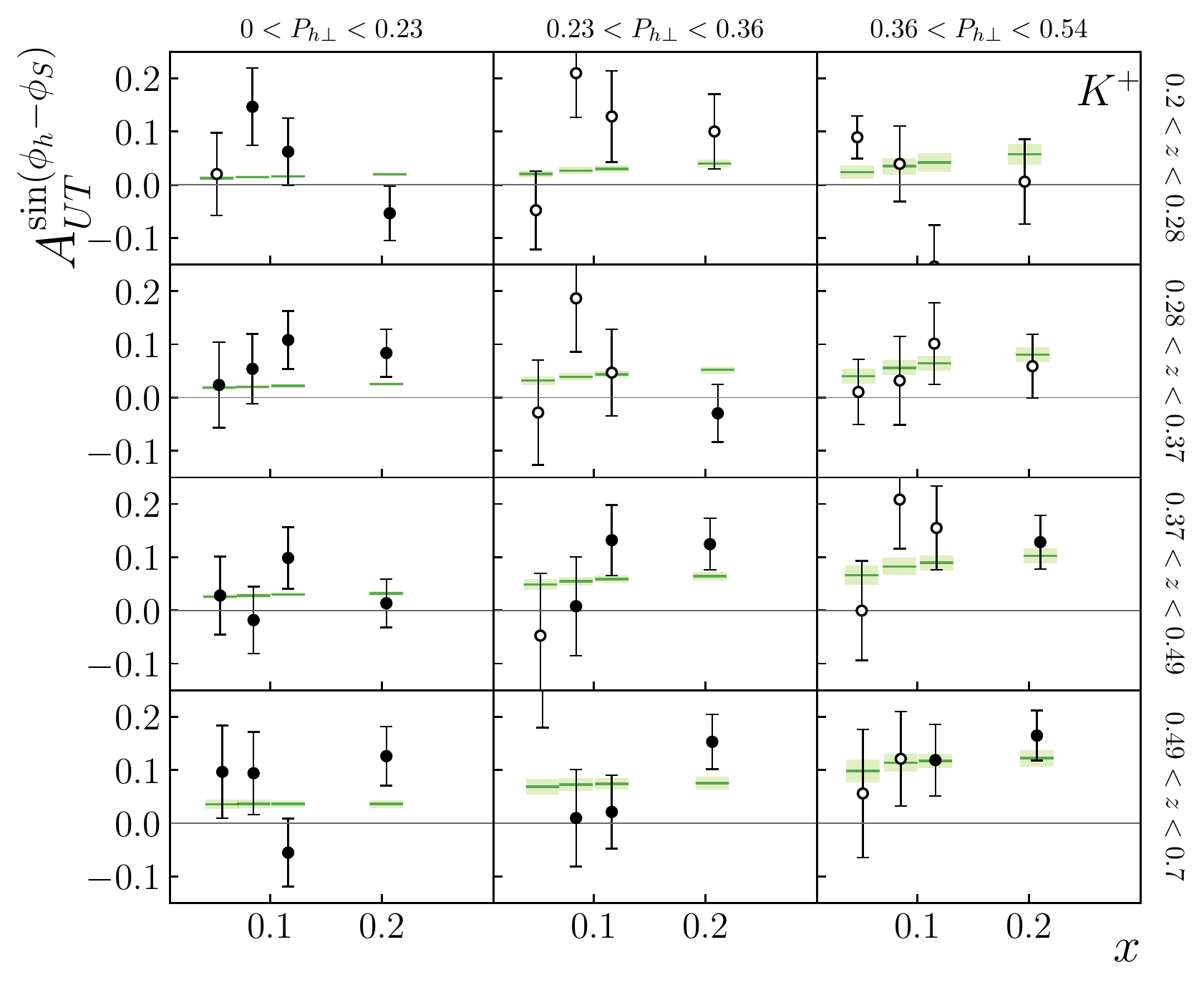}
    \includegraphics[width=0.4\textwidth]{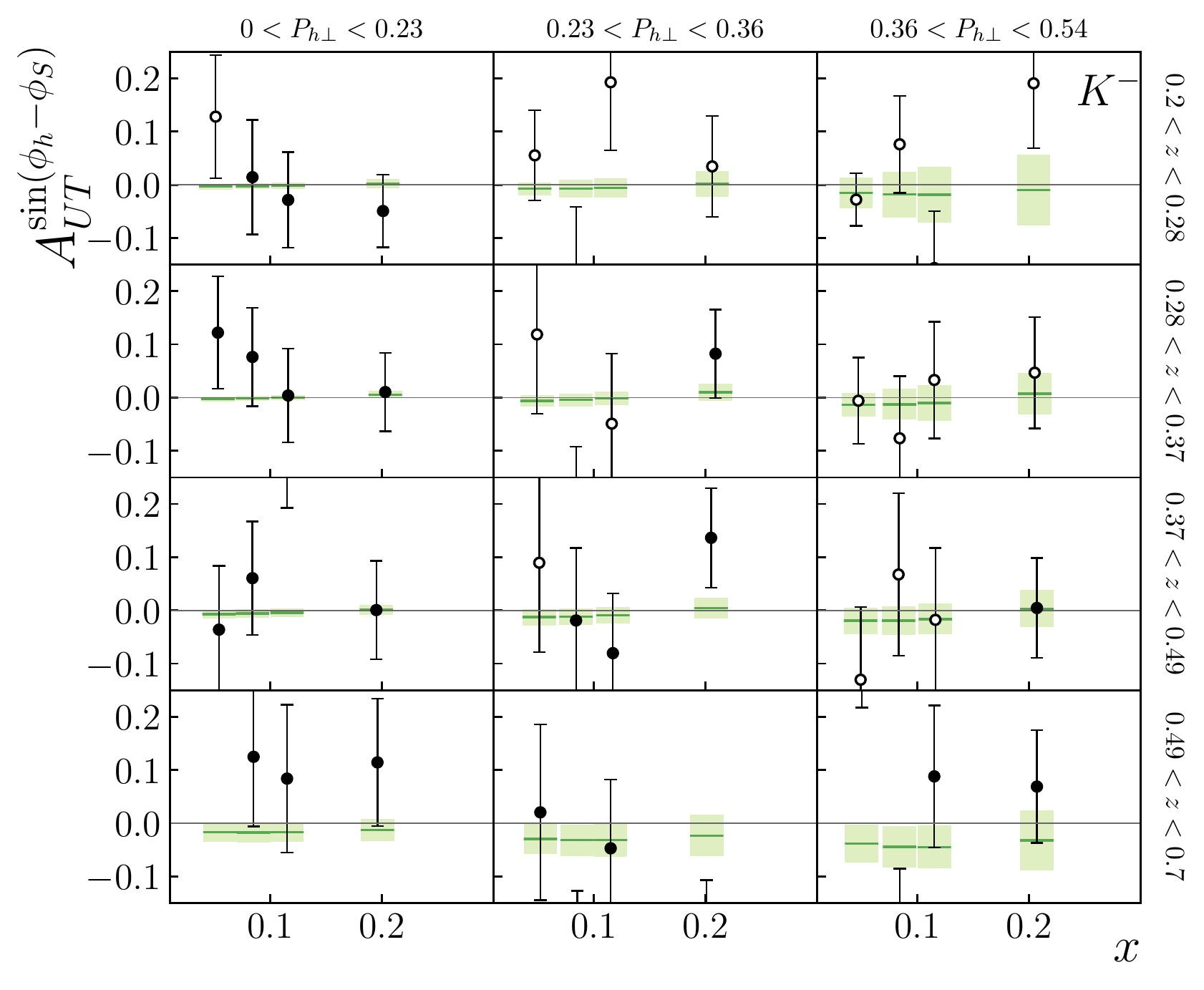}
    \caption{Comparison with HERMES SSA data~\cite{HERMES:2020ifk} from the proton target for $\pi^+$ (upper left), $\pi^-$ (upper right), $K^+$ (lower left), and $K^-$ (lower right) productions. The filled data points are included in the fit, while the open data points are not. The green lines are the central value calculated from the fit and the bands represent the 1 standard deviation of the calculated asymmetries by using 100 replicas.} 
    \label{fig:hermes}
\end{figure*}

\begin{figure*}[htp]
    \centering
    \includegraphics[width=0.8\textwidth]{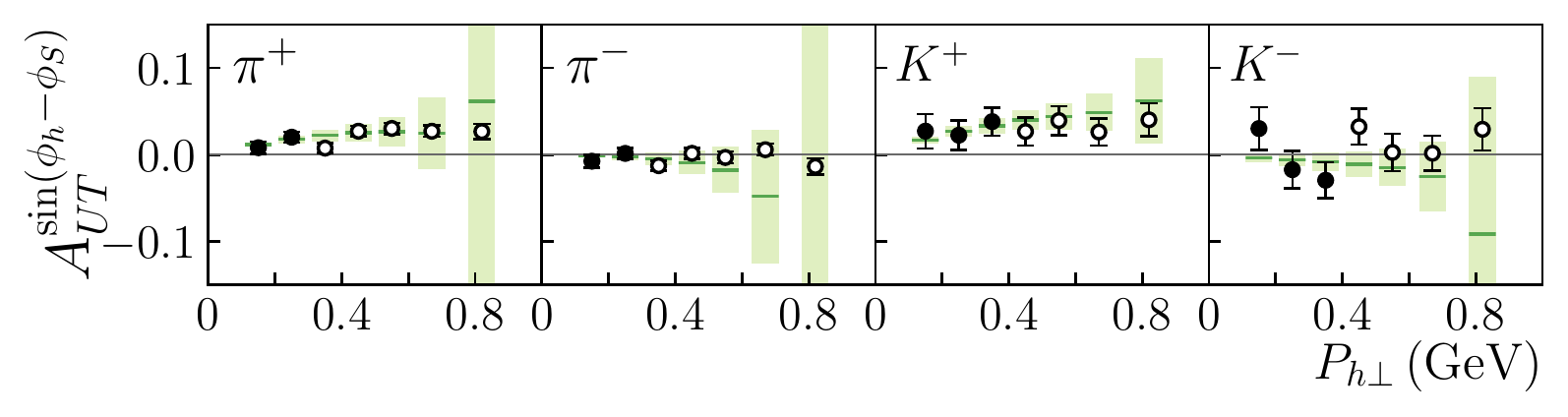}
    \caption{Comparison with COMPASS SSA data~\cite{COMPASS:2014bze} from the proton target for $\pi^+$, $\pi^-$, $K^+$, and $K^-$ productions. The markers and bands have the same meaning as in Fig.~\ref{fig:hermes}.}
    \label{fig:compass_proton}
\end{figure*}

\begin{figure*}[htp]
    \centering
    \includegraphics[width=0.8\textwidth]{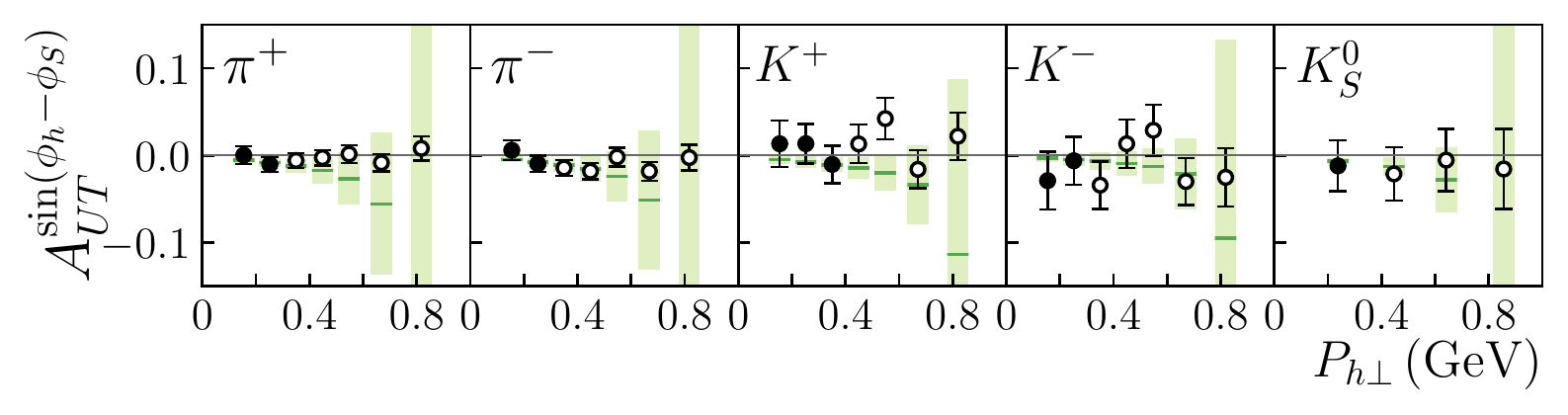}
    \caption{Comparison with COMPASS SSA data~\cite{COMPASS:2008isr} from the deuteron target for $\pi^+$, $\pi^-$, $K^+$,  $K^-$, and $K_S^0$ productions. The markers and bands have the same meaning as in Fig.~\ref{fig:hermes}.}
    \label{fig:compass_deuteron}
\end{figure*}

\begin{figure*}[htp]
    \centering
    \includegraphics[width=0.4\textwidth]{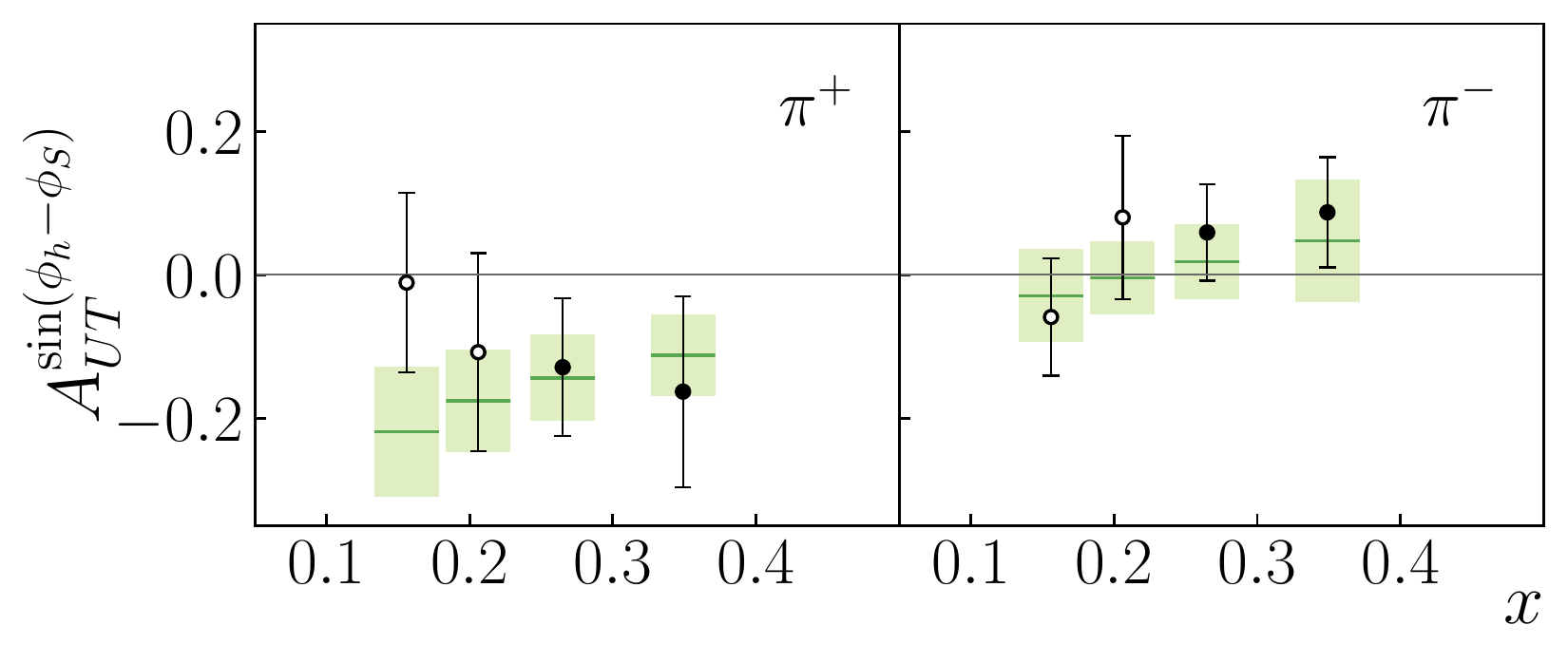}
    \includegraphics[width=0.4\textwidth]{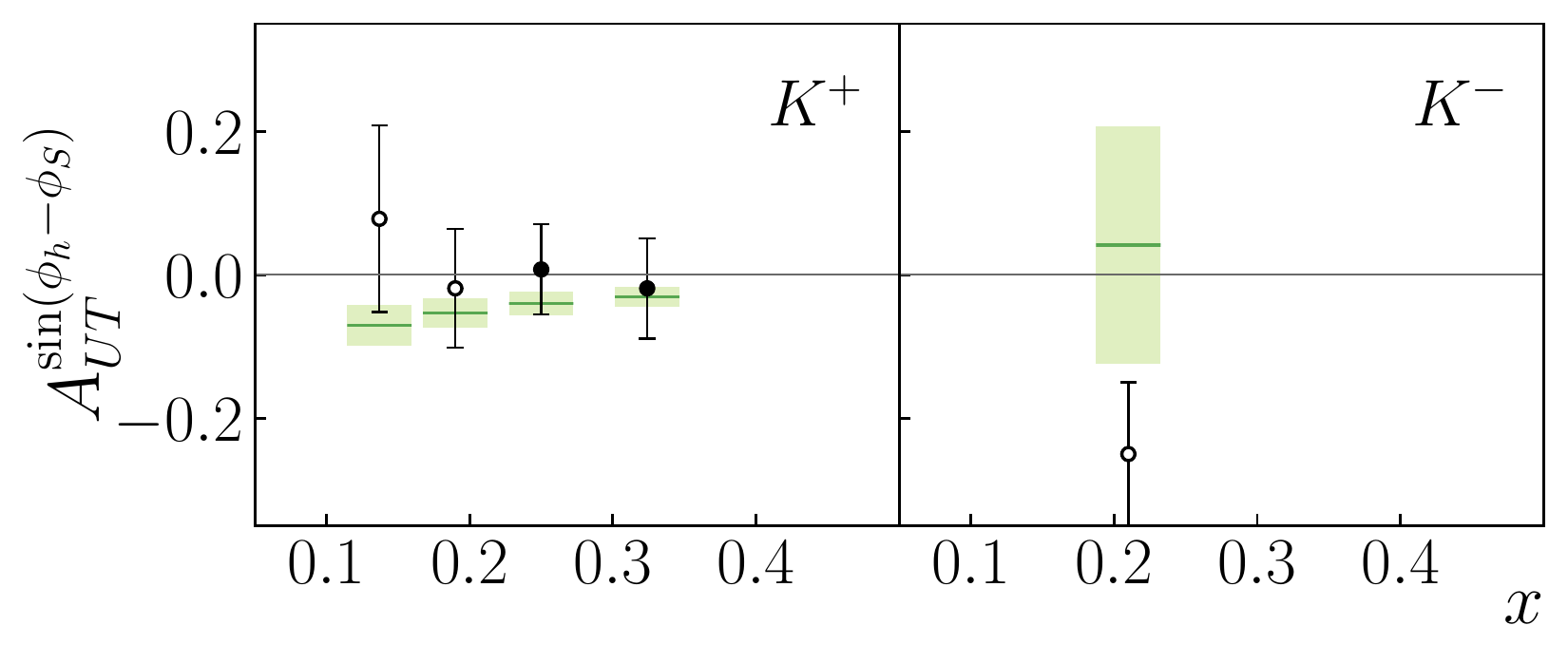}
    \caption{Comparison with JLab SSA data~\cite{JeffersonLabHallA:2011ayy,JeffersonLabHallA:2014yxb} from the $^3\rm He$ target. The asymmetry values for $\pi^+$ and $\pi^-$ productions in the left panel have been effectively converted to those of the neutron, and the asymmetry values for $K^+$ and $K^-$ productions. The markers and bands have the same meaning as in Fig.~\ref{fig:hermes}.}
    \label{fig:jlab}
\end{figure*}

\begin{figure}[htp]
    \centering
    \includegraphics[width=0.98\columnwidth]{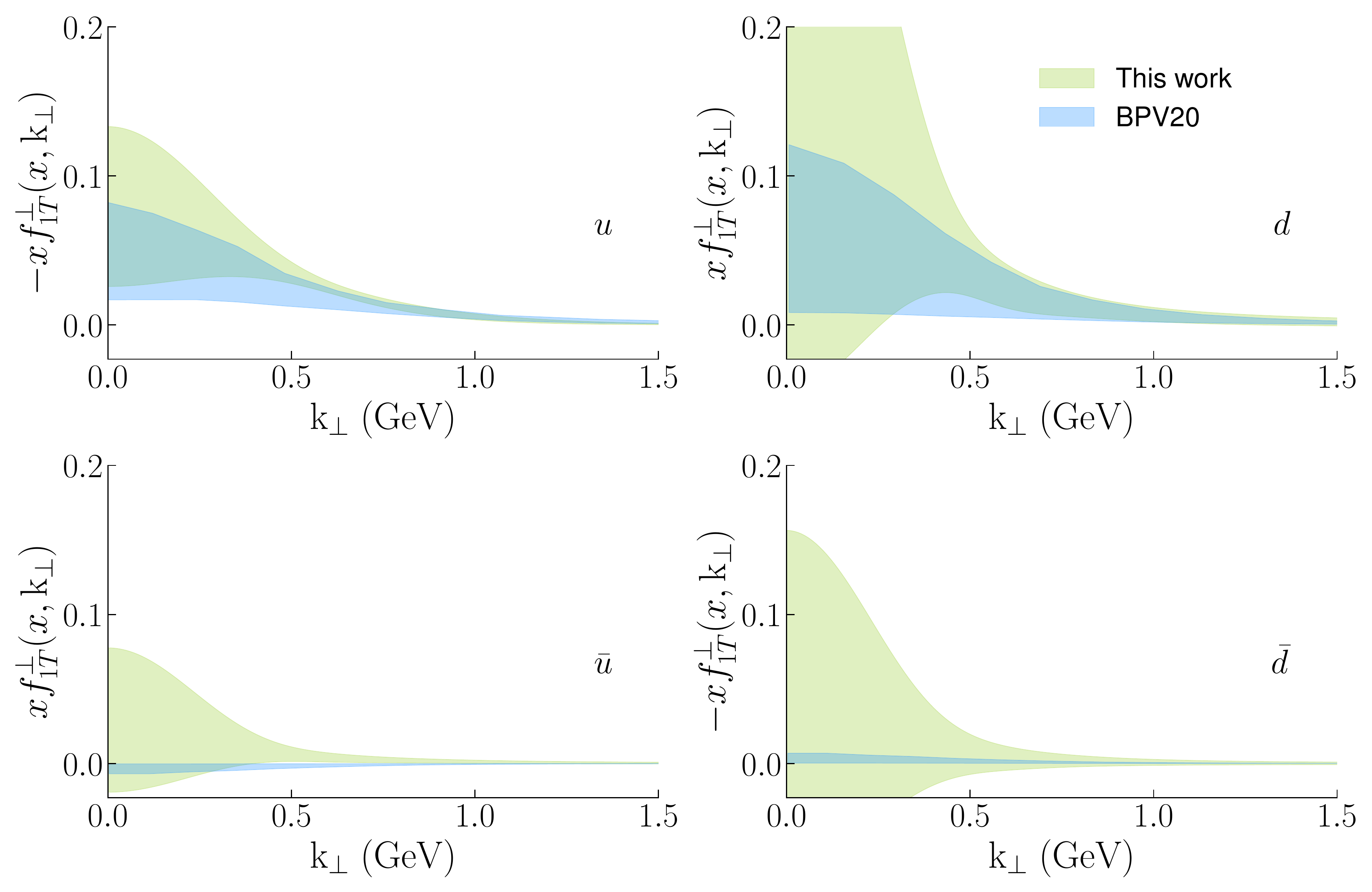}
    \caption{The comparisons between our result and the
    result from Ref. \cite{Bury:2021sue} for    
       Sivers functions of $u, d, \bar{u}, \bar{d}$ quark in the momentum space at $x=0.
    1$ and $Q = 2$ GeV. Within uncertainty, the results are consistent with each other. }
    \label{fig:comparison}
\end{figure}

The central values of the parameters together with their uncertainties are listed in Table~\ref{tab:world_params}. The $\beta$ parameters turn out to be negative for up and down quarks while positive for sea quarks, which gives us some hints that in small-$x$ region the Sivers effect for sea quarks is weaker than that of up and down quarks. On the other hand, existing world data are not
precise enough, and therefore a decisive conclusion will rely on future data from electron-ion
colliders. The comparisons between experimental data and the calculations by using replicas are shown in Figs.~\ref{fig:hermes}-\ref{fig:jlab}, where the filled data points with $\delta < 0.5$ are included in the fit while the open data points with $\delta > 0.5$ are not. The results of the Sivers functions are shown in Fig.~\ref{fig:xf1t_xslices} via slices at various $x$ values.
For better visualization of the $x$ dependence, we also present ${\bf k}_\perp$-integrated distributions of the Sivers function via its zeroth and first transverse momentum moments,

\vspace{-0.5cm}
\begin{align}
    f_{1T}^{\perp (0)}(x)
    &=  \pi \int d {\bf k}_{\perp}^2\, f_{1T}^{\perp}(x, {\bf k}_{\perp}^2),
    \label{eq:0th-moment}
    \\
    f_{1T}^{\perp (1)}(x)
    &= \pi \int d {\bf k}_{\perp}^2\, \frac{{\bf k}_{\perp}^2}{2M^2}f_{1T}^{\perp}(x, {\bf k}_{\perp}^2).
    \label{eq:1st-moment}
\end{align}
Since TMDs are well defined at small transverse momentum and the fit only includes data in the small transverse momentum region, we truncate the integrals at ${\bf k}_\perp^{\rm max} = 0.6\,\rm GeV$. The truncated zeroth and first transverse momentum moments are shown in Fig.~\ref{fig:xf1t0} and Fig.~\ref{fig:xf1t1} respectively. 
Our results are compared with BPV20~\cite{Bury:2021sue} in Fig. \ref{fig:comparison}. Within uncertainty, the results are consistent with each other. However, the uncertainties of sea quark Sivers functions in  Ref.~\cite{Bury:2021sue} are much smaller than our estimations because of strong assumptions in the parametrization. To be more specific, in Ref.~\cite{Bury:2021sue} the intrinsic $k_\perp$ dependence for all the quarks is assumed to be the same, and additionally, the Sivers functions
of sea quarks ($\bar{u}$, $\bar{d}$, $\bar s$, ${s}$)
are assumed to be identical, except for a separate normalization parameter for the $s$ quark.

\begin{table}[htp]
    \centering
    \caption{The values of the parameters from the fit to world SIDIS SSA data. The central values are the average of the results from 100 fits, and the uncertainties are the standard deviations. The values of $r_u$, $r_d$, and $r_{\rm sea}$ are provided in unit of $\rm GeV^2$ and the others are dimensionless. }
    \label{tab:world_params}
    \begin{tabular}{m{0.23\columnwidth}m{0.23\columnwidth}|m{0.23\columnwidth}m{0.23\columnwidth}}
    \hline\hline
    Parameter & Value & Parameter & Value \\
    $r_u$ & $0.08^{+0.04}_{-0.03}$ &
    $N_u$ & $-0.08^{+0.02}_{-0.04}$ \\
    $r_d$ & $0.2^{+0.9}_{-0.2}$ &
    $N_d$ & $1.0^{+2.6}_{-0.5}$ \\
    $r_{\rm sea}$ & $0.2^{+1.5}_{-0.2}$ &
    $N_{\bar{u}}$ & $0.1^{+0.7}_{-0.1}$ \\
    $\beta_u$ & $-0.5^{+0.2}_{-0.2}$ &
    $N_{\bar{d}}$ & $0.0^{+0.9}_{-0.2}$\\
    $\beta_d$ & $-0.97^{+0.12}_{-0.02}$ &
    $\epsilon_u$ & $10^{+7}_{-2}$  \\
    $\beta_{\rm sea}$ & $0.4^{+2.2}_{-0.8}$ &   $\epsilon_d$ & $113^{+215}_{-82}$ \\
    \hline\hline
    \end{tabular}
\end{table}

\begin{table*}
\centering
\caption{The $\chi^2$ values for different datasets. $N$ is the number of
data points for each experimental dataset.}
\label{table:world_chi2}
\begin{tabular*}{0.55\textwidth}{m{0.2\textwidth}m{0.15\textwidth}m{0.2\textwidth}}
\hline\hline
Dataset       & $N$        & $\chi^2/N$   \\ \hline
COMPASS~\cite{COMPASS:2008isr}    &10  &    $1.08_{-0.41}^{+0.52}$   \\

COMPASS~\cite{COMPASS:2014bze}     &10  &  $1.29_{-0.44}^{+0.61}$   \\

HERMES~\cite{HERMES:2020ifk} $\pi^+$  &26  & $1.95_{-0.48}^{+0.48}$         \\
HERMES~\cite{HERMES:2020ifk} $\pi^-$  &26  & $1.83_{-0.37}^{+0.46}$        \\
HERMES~\cite{HERMES:2020ifk} $K^+$  &26  &  $2.23_{-0.48}^{+0.63}$       \\
HERMES~\cite{HERMES:2020ifk} $K^-$  &26  &  $2.35_{-0.48}^{+0.49}$       \\

JLab~\cite{JeffersonLabHallA:2011ayy}\cite{JeffersonLabHallA:2014yxb}    & 6  &  $1.03_{-0.46}^{+0.75}$   \\ 
\hline
total                          &130      &$1.90_{-0.18}^{+0.18}$  \\
\hline \hline
\end{tabular*}
\end{table*}


\subsection{EicC projections of the Sivers function}
\label{sec:fiteicc}

The SIDIS events are generated at the vertex level using a Monte Carlo event generator~\cite{generator}, which has been adopted in the simulation of several JLab12 SIDIS experiments. It allows us to run in collider mode. According to the conceptual design of EicC, the electron beam energy is set at $3.5\,\rm GeV$, the proton beam energy is set at $20\,\rm GeV$, and the $^3$He beam energy is set at $40\,\rm GeV$. The unpolarized SIDIS differential cross section used in the generator is based on global fit to HERMES and COMPASS multiplicity data. A similar cross section parametrization is also adopted in {\tt SIDIS-RC EvGen}~\cite{rcgenerator}, a recently released generator for radiative correction studies in the SIDIS process. For acceptance reasons, we require the electron momentum greater than $0.35\,\rm GeV$ and the hadron momentum greater than $0.30\,\rm GeV$. The pseudorapidity $\eta$ is restricted to $-3.5 < \eta < 3.5$ for both the electron and the hadron. Full azimuthal angle coverage in the laboratory frame is assumed. To have full flavor separation of all light quarks, we include both charged pion, $\pi^\pm$, and charge kaon, $K^\pm$, productions in SIDIS. For the particle identification (PID) of the final electron, pion, and kaon, we apply a PID acceptance cut on the maximum particle momentum, $p_{\rm max}$, in different pseudorapidity regions,
\begin{center}
\begin{tabular}{lccc}
\hline\hline
$\eta$  & ~~~~$[-3.5,-1]$~~~~ & ~~~~$(-1,1]$~~~~ & ~~~~$(1,3.5]$~~~~ \\
$p_{\rm max}$ & $4\,\rm GeV$ & $6\,\rm GeV$ & $15\,\rm GeV$\\
\hline\hline
\end{tabular}
\end{center}
where the initial ion beam is defined along the positive direction. We further impose the physical cuts, $Q^2 > 1\,\rm GeV^2$, $0.3 < z < 0.7$, $W > 5\,\rm GeV$, and $W' > 2\,\rm GeV$, to select events in the deep inelastic region and to exclude resonance regions. 

\begin{figure}[htp]
    \centering
    \includegraphics[width=0.98\columnwidth]{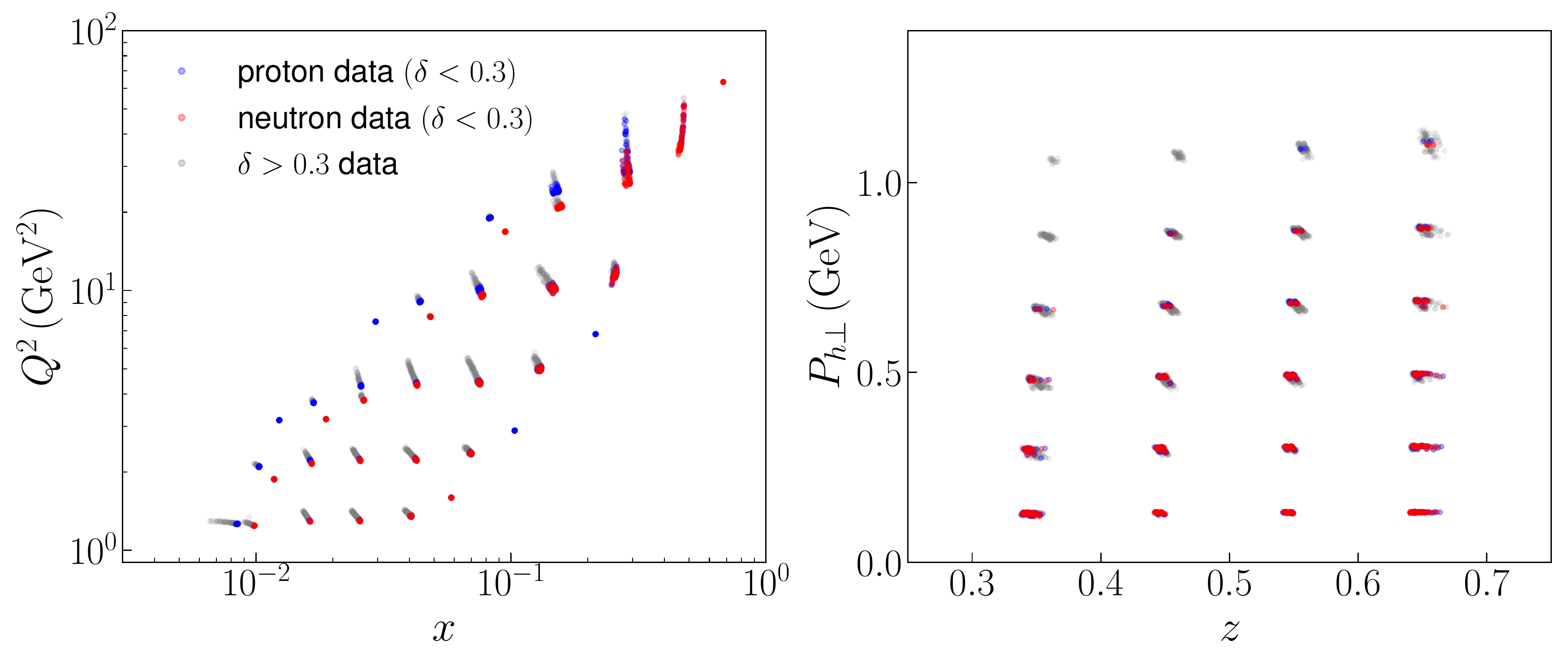}
    \caption{Kinematic distributions of the EicC pseudodata in $x-Q^2$ (left) and $z-P_{h\perp}$ (right) planes. Each bin is plotted as a point at the bin center kinematic values. Kinematics cuts, $Q^2 > 1\,\rm GeV^2$, $W>5\,\rm GeV$, $W'>2\,\rm GeV$, and $0.3<z<0.7$ have been applied. The blue points are the proton data with $\delta<0.3$, the red points are the neutron data with $\delta<0.3$, and the gray points are the data with $\delta > 0.3$. }
    \label{fig:eicc_kin}
\end{figure}

We estimate the statistics by assuming $50\,\rm fb^{-1}$ for $ep$ collisions and $50\,\rm fb^{-1}$ for $e {^3}\rm He$ collisions. Such accumulated luminosities can be achieved with about one year running according to the designed instantaneous luminosity $2\times 10^{33}\,\rm cm^{-2} s^{-1}$. Keeping the statistical uncertainty at $10^{-3}$ level, we obtain 13545 data points in four-dimensional bins in $x$, $Q^2$, $z$, and $P_{h\perp}$. Not only the precision of the EicC pseudodata is much higher than existing world data, but also the amount of data points is about 10 times more. It allows us to apply more strict kinematic cuts for a cleaner selection of data in the TMD region. In this study, we choose $\delta < 0.3$, and 4983 EicC pseudodata points are selected. The distributions of the EicC pseudodata are shown in Fig.~\ref{fig:eicc_kin}, where the colored points are selected in the fit while the gray ones are excluded. Since the energies per nucleon are different for the proton beam and the $^3\rm He$ ion beam, the kinematics of the proton data and the neutron data are slightly different but still overlap in a wide range.

\begin{figure*}[htp]
    \centering
    \includegraphics[width=0.27\textwidth,page=1]{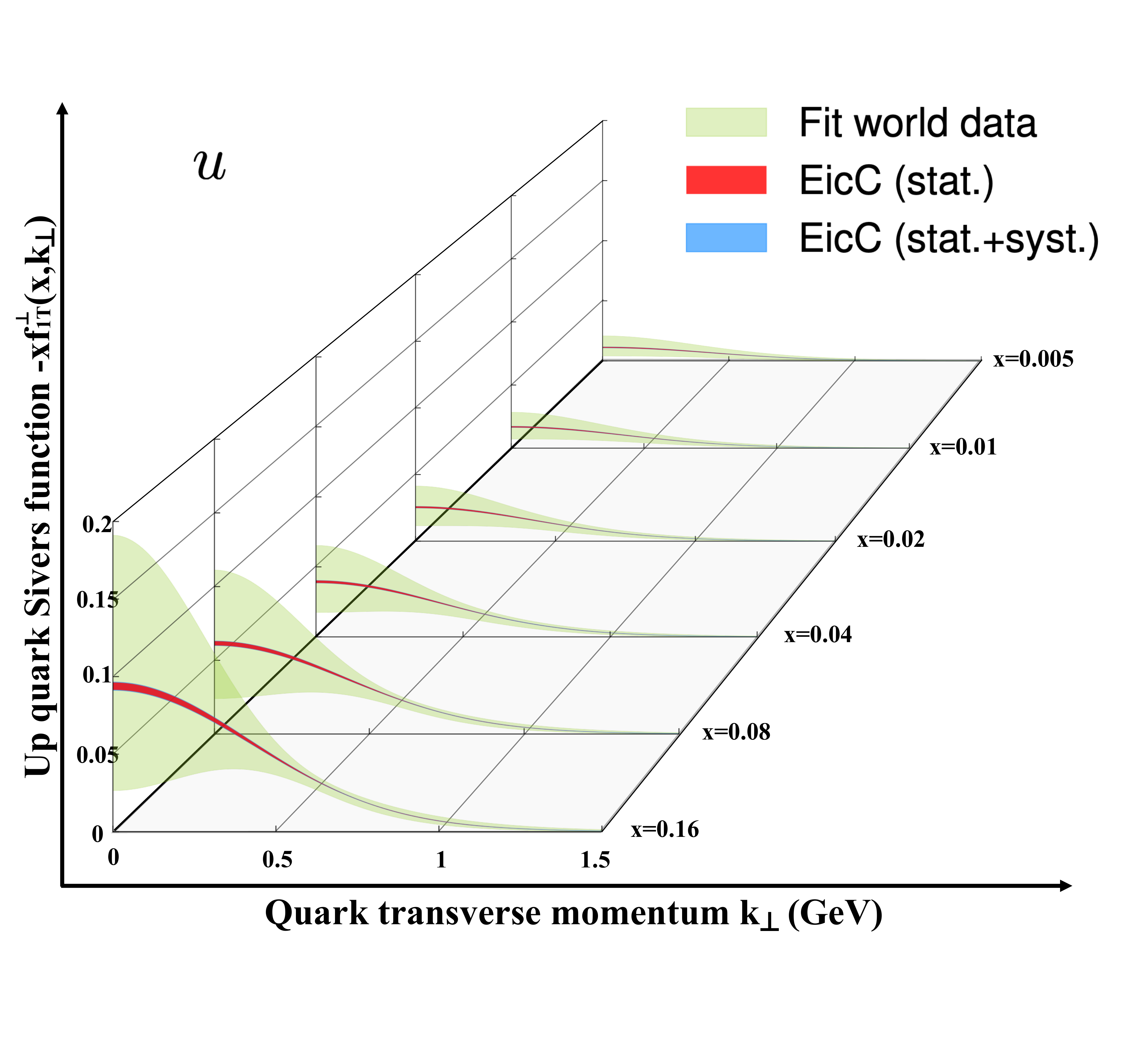}
    \includegraphics[width=0.27\textwidth,page=2]{sivers.pdf}
    \includegraphics[width=0.27\textwidth,page=3]{sivers.pdf}
    \includegraphics[width=0.27\textwidth,page=4]{sivers.pdf}
    \includegraphics[width=0.27\textwidth,page=5]{sivers.pdf}
    \includegraphics[width=0.27\textwidth,page=6]{sivers.pdf}
    \caption{The transverse momentum distribution of the Sivers functions at different $x$ values. The green bands represent the uncertainties of the fit to world SIDIS data, the red bands represent the EicC projections with only statistical uncertainties, and the blue bands represent the EicC projections including systematic uncertainties as described in the text.}
    \label{fig:xf1t_xslices}
\end{figure*}

\begin{figure*}[htp]
\includegraphics[width=0.8\textwidth]{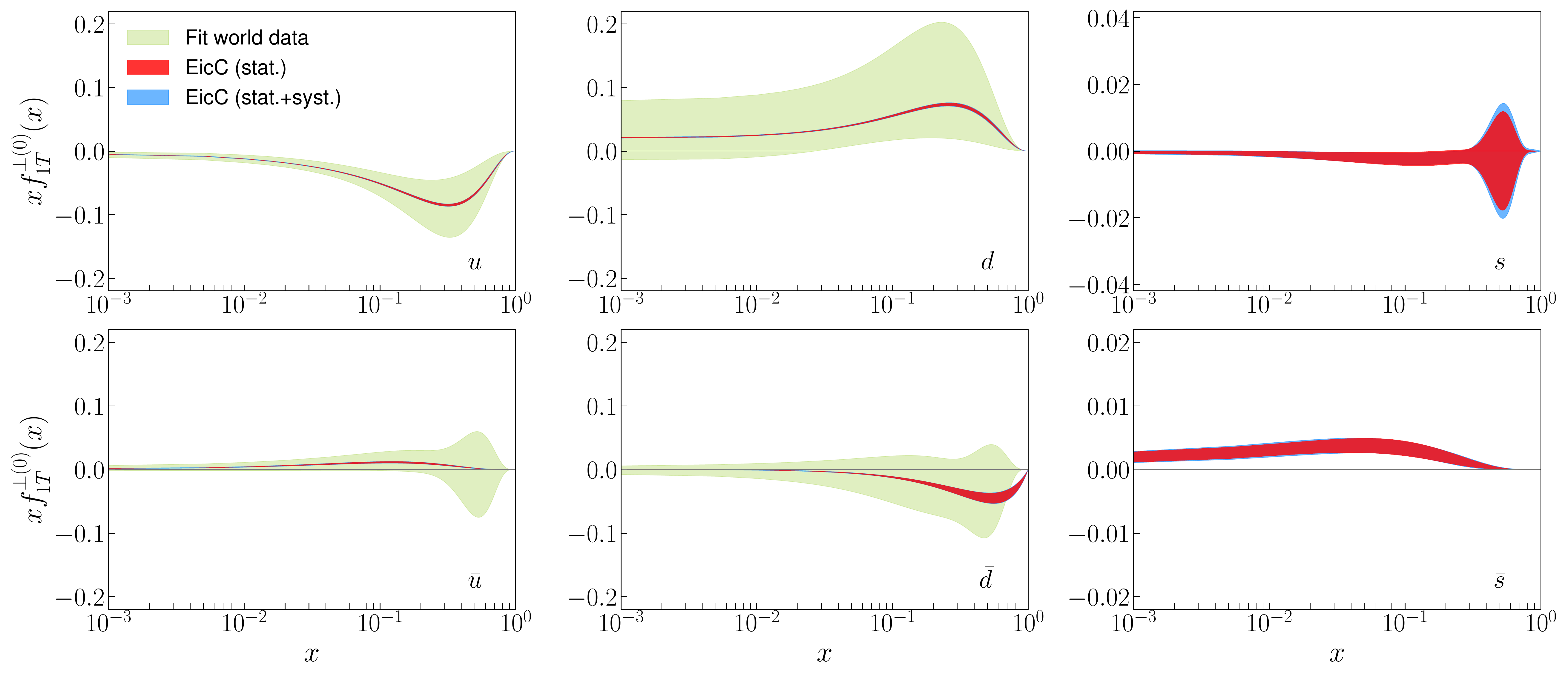}
\caption{The zeroth transverse moment of the Sivers functions as defined in Eq.~\eqref{eq:0th-moment} with the ${\bf k}_\perp$ integral truncated at $0.6\,\rm GeV$. The green bands represent the uncertainties of the fit to world SIDIS data, the red bands represent the EicC projections with only statistical uncertainties, and the blue bands represent the EicC projections including systematic uncertainties as described in the text.}
\label{fig:xf1t0}
\end{figure*}

\begin{figure*}[htp]
\includegraphics[width=0.8\textwidth]{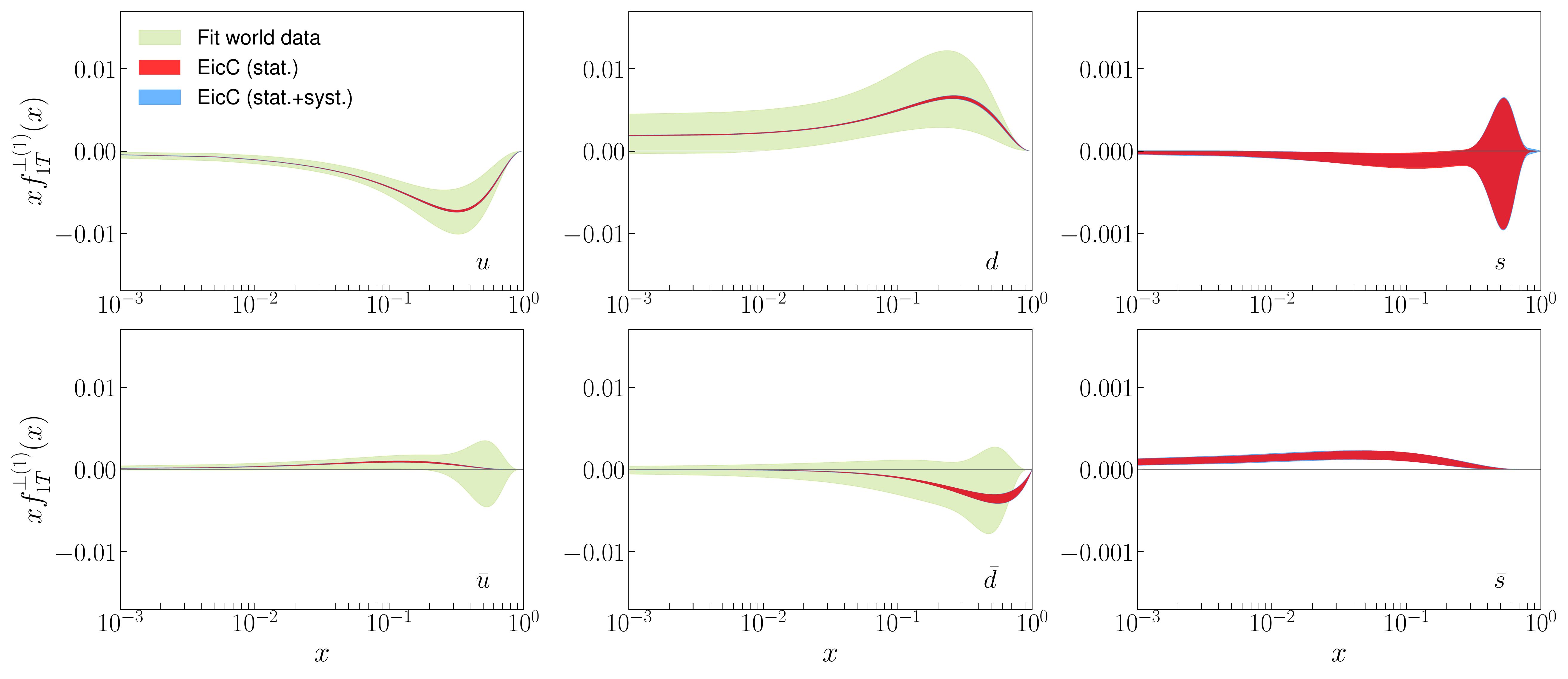}
\caption{The first transverse moment of the Sivers functions as defined in Eq.~\eqref{eq:1st-moment} with the ${\bf k}_\perp$ integral truncated at $0.6\,\rm GeV$. The green bands represent the uncertainties of the fit to world SIDIS data, the red bands represent the EicC projections with only statistical uncertainties, and the blue bands represent the EicC projections including systematic uncertainties as described in the text.}
\label{fig:xf1t1}
\end{figure*}

The SSA values of the pseudodata are calculated with the central value of the result from the fit to the world data. Since a realistic estimation of systematic uncertainties is only possible when the detailed designs of detectors are available, we only consider some expected dominant sources of systematic uncertainties. For the proton data, we assign $3\%$ relative uncertainty to account for the polarization of the proton beam, and for the neutron data, we assign $5\%$ relative uncertainty to account for the polarization of the $^3\rm He$ ion beam and the nuclear effect. Total uncertainties are evaluated via the quadrature combination of statistical uncertainties and systematic uncertainties. The precise EicC data with wide kinematics coverage allows us to adopt a more flexible parametrization of the Sivers functions. Therefore, we remove the artificial assumptions in Eq.~\eqref{eq:param-constraints}, while still keep $\epsilon_{\bar u}=\epsilon_{\bar d}=\epsilon_{s}=\epsilon_{\bar s}=0$, and then we have 26 free parameters, as listed in Table~\ref{tab:eicc_params}. To estimate the impact of the EicC on the extraction of the Sivers function, we perform a simultaneous fit to the world data and the EicC pseudodata as described above. Following the same procedure, 100 replicas are created by randomly shifting the values according to the simulated statistical uncertainty. The fit reaches $\chi^2/N = 1.15$ for only statistical uncertainties and $\chi^2/N = 1.13$ for both statistical and systematic uncertainties. The average values of the parameters and their uncertainties are provided in Table~\ref{tab:eicc_params}. The results of the EicC projection of the Sivers functions are shown in Fig.~\ref{fig:xf1t_xslices} via slices at various $x$ values, in Fig.~\ref{fig:xf1t0} via the truncated zeroth transverse momentum moment, and in Fig.~\ref{fig:xf1t1} via the truncated first transverse momentum moment in comparison with the results of the fit to existing world data.

\begin{table*}[htp]
    \centering
    \caption{The parameters from the fit to world SIDIS data and EicC pseudodata. The central values are the average of the results from 100 fits, and the uncertainties are the standard deviations. The values of $r_u$, $r_d$, $r_s$, $r_{\bar u}$, $r_{\bar d}$, and $r_{\bar s}$ are provided in unit of $\rm GeV^2$ and the others are dimensionless. The ``Stat.” column means that only statistical uncertainties of EicC pseudodata are considered in the fit, while ``Stat. + Syst.” column means that both statistical and systematic uncertainties of EicC pseudodata are
included in the fit. }
    \label{tab:eicc_params}
    \begin{tabular}{m{0.14\textwidth}m{0.16\textwidth}m{0.16\textwidth}|m{0.14\textwidth}m{0.16\textwidth}m{0.16\textwidth}}
    \hline\hline
    Parameter & Stat. & Stat.+Syst. & Parameter & Stat. & Stat.+Syst. \\
    $r_u$ & $0.068^{+0.002}_{-0.001}$ & $0.067^{+0.002}_{-0.002}$ &
    $N_u$ & $-0.075^{+0.001}_{-0.001}$ & $-0.075^{+0.001}_{-0.001}$ \\
    $r_d$ & $0.092^{+0.003}_{-0.003}$ & $0.091^{+0.003}_{-0.003}$&
    $N_d$ & $0.72^{+0.02}_{-0.02}$ & $0.72^{+0.02}_{-0.02}$\\
    $r_{s}$ & $0.005^{+0.044}_{-0.005}$ & $0.009^{+0.054}_{-0.009}$ &
    $N_s$ & $-0.001^{+0.001}_{-0.001}$ & $-0.001^{+0.001}_{-0.001}$  \\
    $r_{\bar u}$ & $0.065^{+0.011}_{-0.008}$ & $0.064^{+0.012}_{-0.009}$ &
    $N_{\bar{u}}$ & $0.012^{+0.001}_{-0.001}$ & $0.012^{+0.001}_{-0.001}$ \\
    $r_{\bar d}$ & $0.044^{+0.008}_{-0.006}$ & $0.044^{+0.007}_{-0.007}$ &
    $N_{\bar{d}}$ & $-0.016^{+0.001}_{-0.001}$ & $-0.016^{+0.001}_{-0.001}$ \\
    $r_{\bar s}$ & $1.1^{+6.7}_{-1.0}\times 10^{-8}$ & $1.5^{+12.2}_{-1.4}\times 10^{-8}$ &
    $N_{\bar{s}}$ & $0.002^{+0.001}_{-0.001}$ & $0.001^{+0.001}_{-0.001}$ \\
    $\beta_u$ & $-0.44^{+0.02}_{-0.03}$ & $-0.44^{+0.02}_{-0.03}$ &
    $\alpha_u$ & $2.57^{+0.04}_{-0.04}$ & $2.55^{+0.07}_{-0.05}$ \\
    $\beta_d$ & $-0.9840^{+0.0003}_{-0.0003}$ & $-0.9840^{+0.0003}_{-0.0004}$ &
    $\alpha_d$ & $2.52^{+0.06}_{-0.05}$ & $2.56^{+0.08}_{-0.06}$ \\
    $\beta_s$ & $0.2^{+3.8}_{-0.6}$ & $0.2^{+4.1}_{-0.6}$ &
    $\alpha_s$ & $6^{+4}_{-2}$ & $6^{+4}_{-2}$ \\
    $\beta_{\bar u}$ & $-0.37^{+0.04}_{-0.04}$ & $-0.38^{+0.05}_{-0.05}$ &
    $\alpha_{\bar u}$ & $4.60^{+0.36}_{-0.27}$ & $4.53^{+0.39}_{-0.39}$ \\
    $\beta_{\bar d}$ & $0.17^{+0.08}_{-0.07}$ & $0.16^{+0.10}_{-0.08}$ &
    $\alpha_{\bar d}$ & $1.003^{+0.008}_{-0.002}$ & $1.004^{+0.006}_{-0.003}$ \\
    $\beta_{\bar s}$ & $-0.7^{+0.1}_{-0.1}$ & $-0.7^{+0.2}_{-0.1}$ &
    $\alpha_{\bar s}$ & $7^{+3}_{-1}$ & $7^{+3}_{-2}$ \\
    $\epsilon_u$ & $5.5^{+0.9}_{-0.5}$ & $5.5^{+1.2}_{-0.6}$ & 
    $\epsilon_d$ & $24^{+11}_{-11}$ & $25^{+13}_{-12}$ \\
    \hline\hline
    \end{tabular}
\end{table*}

\section{Summary}
\label{sec:summary}

In this paper, we present a quantitative assessment of the impact of EicC SIDIS program on the determination of TMDs. Taking the Sivers function as an example, we perform a global fit of the Sivers asymmetry data in SIDIS at small transverse momentum, including the TMD evolution at the NNLL accuracy. The impact of EicC is studied by adding the EicC pseudodata. In this study, both  statistical uncertainties and dominant systematic uncertainties are taken into account for the EicC pseudodata,
while complete detailed systematic uncertainty studies are left for the future when the detector design is ready.
It has been demonstrated that the Sivers
functions can be precisely determined for various quark flavors, and particularly the sea quark distributions, including the strange and antistrange, can be extracted at high precision with the future EicC SIDIS data. 

Once the precise data are available from EicC, one will be able to have less biased extractions of the Sivers functions by using much more flexible parametrizations. Besides, in the EicC era, one can have a cleaner selection of data for TMD studies, {\it e.g.}, by applying a more strict requirement on $\delta \equiv |P_{h\perp}|/(z Q)$ to restrict data in the low transverse momentum region and higher $W$ and $W'$ cuts to avoid the resonance region. It is important to remark that both polarized electron-proton 
and electron-$^{3}$He
data are necessary for a complete flavor separation. To fully explore the potential of $^3$He as an effective
neutron source, detailed nuclear effect corrections should be further investigated both
experimentally and theoretically
in the future, since it is the dominant source of systematic uncertainty in reality by using
$^3$He data.

In principle, the EicC enables us to measure all 18 TMDs-related structure functions in SIDIS via the combination of different electron and ion beam polarization configurations and the separation of different azimuthal modulations. The study of the Sivers function as presented in this paper can be extended to other TMDs. Multi dimensional binning on $x$, $Q^2$, $z$ and $p_{h\perp}$ will be available for the spin asymmetry measurements, and the coverage of $x$ by EicC can reach down to about $0.005$. Given the existing fixed-target
experiments covering the low-$Q^2$ and high-$x$ region and the Electron-Ion Collider to be built at BNL in US (US-EIC) reaching much lower $x$ values, EicC will fill the kinematics gap between the coverage between the JLab-12GeV program and the US-EIC. Combining
the measurements at all these facilities, we will be able to have a complete physical picture of the three-dimensional structures of the nucleon with systematically controllable uncertainties. Therefore, EicC will play an important role in the understanding of nucleon spin structures with its unique significance for sea quarks.

\acknowledgments{
This work is supported in part by the Strategic Priority Research Program of the Chinese Academy of Sciences under grant number XDB34000000,
the Guangdong Provincial Key Laboratory of Nuclear Science with No. 2019B121203010, and the National Natural Science Foundation of China under Contract No. 12175117. 
P. Sun is supported by the Natural Science Foundation of China under grant No. 11975127
and No. 12061131006. The authors acknowledge the computing resource available at the Southern Nuclear Science Computing Center.
}

\newpage

\appendix
\vspace{1in}
\section{EXPRESSION OF MATCHING FUNCTION}
\label{sec:appendix}

For TMD PDFs, the coefficient function $C$ up to NLO is~\cite{Scimemi:2019cmh}

\vspace{-0.5cm}
\begin{align}\label{eq:cpdf}
     &C_{f\gets f'}(x, b, \mu)
     =\delta(1-x)\delta_{ff'}\notag\\
     &\quad\quad\quad\quad
     +a_s{(\mu)}\Big(-\textbf{L}_{\mu} P^{(1)}_{f\gets f'}+C^{(1,0)}_{f\gets f'}  \Big) \ ,
\end{align}
where

\vspace{-0.5cm}
\begin{align}
     C^{(1,0)}_{q\gets q'}(x)
     &=C_F\Big[2(1-x)-\delta(1-x)\frac{\pi^2}{6}  \Big]  \delta_{q q'} ,\\
     C^{(1,0)}_{q\gets g}(x)&=2x(1-x) ,\\
     P^{(1)}_{q\gets q'}(x)&=2C_F \Big[\frac{2}{(1-x)_+}-1-x+\frac{3}{2}\delta(1-x) \Big] \delta_{q q'} , \\
     P^{(1)}_{q\gets g}(x)
     &=1-2x+2x^2 . 
\end{align}

For TMD FFs, the matching coefficient $\mathbb{C}$ up to NLO follows the same pattern as in Eq.~\eqref{eq:cpdf} with the replacement of the PDF DGLAP kernels $P^{(1)}_{f\gets f'}(x)$ by the FF DGLAP kernels~\cite{Stratmann:1996hn},

\vspace{-0.5cm}
\begin{align}
     {\mathbb P}^{(1)}_{q\to q'}(z)=&\frac{2C_F}{z^2}\Big(\frac{1+z^2}{1-z}\Big)_+  \delta_{q q'},\\
     {\mathbb P}^{(1)}_{q\to g}(z)=&\frac{2C_F}{z^2}\frac{1+(1-z)^2}{z} ,
\end{align}
and the replacement of $C^{(1,0)}_{f\gets f'}(x)$ by~\cite{Scimemi:2019cmh} 

\vspace{-0.5cm}
\begin{align}
     \mathbb{C}^{(1,0)}_{q\to q'}(z)=&\frac{C_F}{z^2}\Big[2(1-z)+\frac{4(1+z^2)\ln{z}}{1-z}\notag\\&
     -\delta(1-z)\frac{\pi^2}{6}  \Big]\delta_{q q'} ,\\
     \mathbb{C}^{(1,0)}_{q\to g}(z)=&\frac{2C_F}{z^2}\Big[z+2\big(1+(1-z)^2\big)\frac{\ln{z}}{z}     \Big].
\end{align}
The ``+'' prescription is defined as

\vspace{-0.5cm}
\begin{align}
   &\int_{x_0}^1 dx\, [g(x)]_+ f(x)\notag\\
   &=\int_0^1 dx\, g(x)[f(x)\Theta(x-x_0)-f(1)],
\end{align}
where $\Theta(x-x_0)$ is the Heaviside step function.

\section{ESTIMATION OF UNCERTAINTIES BY USING REPLICAS}
\label{sec:appendix_2}
We randomly shift the central values of the data points by Gaussian distributions with the Gaussian widths to be the experimental uncertainties. Repeating the process, we get 100 datasets $y=\{y_1, y_2, \cdots, y_i, \cdots, y_{100} \}$. Then each dataset is fitted to 
yield the best estimation of various parameters, and hence a replica is created. Afterward, one can calculate the set $C=\{\chi_1^2/N, \chi_2^2/N, \cdots, \chi_i^2/N, \cdots, \chi_{100}^2/N\}$, where $N$ is the total number
of experimental data points. The central value, $\chi^2/N$, is the mean value of the set $C$. 
In principle, $C$ could be any physical quantity, for example, the Sivers function or the asymmetry $A_{UT}^{\sin(\phi_h - \phi_S)}$. 
In order to calculate the
 upper and lower uncertainties of $\chi^2/N$, we define
$C=\{C_{up}, C_{lo}\}$,
where
$C_{up}=\{\chi_1^2/N, \chi_2^2/N, \cdots, \chi_i^2/N, \cdots, \chi_{k}^2/N   \}$
for $\chi_i^2/N>\chi^2/N$ and 
$C_{lo}=\{\chi_1^2/N, \chi_2^2/N, \cdots, \chi_i^2/N, \cdots, \chi_{100-k}^2/N   \}$ 
for $\chi_i^2/N<\chi^2/N$.
Then we can calculate the upper and lower uncertainties by 
\begin{align}
     \delta^+&=\sqrt{\sum_1^k\frac{ (\chi^2_i/N-\chi^2/N)^2}{k}}
\end{align}
for $\chi^2_i/N$ in $C_{up}$, and
\begin{align}
     \delta^-&=\sqrt{\sum_1^{100-k}\frac{ (\chi^2_i/N-\chi^2/N)^2}{100-k}}
\end{align}
for $\chi^2_i/N$ in $C_{lo}$.


\end{document}